\documentclass[a4paper,fleqn,traditabstract]{aa}
\usepackage{natbib,amssymb,amsmath,url,booktabs,array}
\usepackage{epstopdf}
\usepackage{graphicx}
\usepackage[version=3]{mhchem}
\usepackage[normalem]{ulem}
\DeclareGraphicsExtensions{.pdf,.png,.eps,.gif,.ps}
\graphicspath{  {./fig/} }

\newcommand{\kms}{\ensuremath{\,\text{km~s}^{-1}}}
\newcommand{\s}{\ensuremath{\,\text{s}^{-1}}}

\newcommand{\cmm}{\ensuremath{\,\text{cm}^{-2}}}
\newcommand{\cmmm}{\ensuremath{\,\text{cm}^{-3}}}
\newcommand{\cmmmp}{\ensuremath{\,\text{cm}^{3}}}

\newcommand{\hcop}{\ensuremath{\text{HCO}^+}}
\newcommand{\htrcop}{\ensuremath{\text{H}^{13}\text{CO}^+}}
\newcommand{\dcop}{\ensuremath{\text{DCO}^+}}
\newcommand{\co}{\ensuremath{\text{CO}}}

\newcommand{\thco}{\ensuremath{^{13}\text{CO}}}
\newcommand{\cdo}{\ensuremath{\text{C}^{18}\text{O}}}
\newcommand{\cseo}{\ensuremath{\text{C}^{17}\text{O}}}
\newcommand{\docseo}{\ensuremath{^{12}\text{C}^{16}\text{O}}}

\newcommand{\h}{\ensuremath{\text{H}}}
\renewcommand{\d}{\ensuremath{\text{D}}}

\newcommand{\hh}{\ensuremath{{\rm H_2}}}
\newcommand{\hhp}{\ensuremath{\text{H}_2^+}}
\newcommand{\hd}{\ensuremath{\text{HD}}}

\newcommand{\htrp}{\ensuremath{\text{H}_3^+}}
\newcommand{\hhdp}{\ensuremath{\text{H}_2\text{D}^+}}
\newcommand{\doc}{\ensuremath{^{12}\text{C}}}
\newcommand{\trc}{\ensuremath{^{13}\text{C}}}
\newcommand{\huo}{\ensuremath{^{18}\text{O}}}
\newcommand{\seo}{\ensuremath{^{16}\text{O}}}
\newcommand{\e}{\ensuremath{\text{e}^-}}
\newcommand{\xe}{\ensuremath{x_\text{e}}}
\newcommand{\ceq}[2]{\stackrel {#1}{\underset {#2}{\rightleftharpoons}}}
\newcommand{\carrow}[1]{\stackrel {#1}{\rightarrow}}
\def\tkin{\ensuremath{T_{\rm kin}}}

\def \elec{\ensuremath{\ce{e-}}}
\def \xe{\ensuremath{x_e}}
\def \nh{\ensuremath{n_{\rm H}}}

\def \dix#1{\ensuremath{{\,\rm 10^{#1}}}}
\def \tdix#1{\ensuremath{{\,\times 10^{#1}}}}
\def \rd{\ensuremath{{R_D}}}

\newcounter{exno}
\newenvironment{network}
{
\begin{tabular}{>{\refstepcounter{exno}no. \theexno}lrcll}
}
{
\end{tabular}
}

\newenvironment{mytablecaption}
{
  \begin{footnotesize}
    \textit{Note} - 
  }
  {
  \end{footnotesize}
}

\begin{document}
\setlength{\mathindent}{0pt}

\title{Cosmic ray induced ionisation of a molecular cloud shocked by
  the W28 supernova remnant}

\author{
S. Vaupr\'e \inst{\ref{inst1}} 
\and P. Hily-Blant\inst{\ref{inst1}} 
\and C. Ceccarelli\inst{\ref{inst1}} 
\and G. Dubus\inst{\ref{inst1}}
\and S. Gabici\inst{\ref{inst2}}
\and T. Montmerle\inst{\ref{inst3}}
}

\institute{
Univ. Grenoble Alpes/CNRS, IPAG, F-38000 Grenoble, France;\label{inst1}
  \email{solenn.vaupre@obs.ujf-grenoble.fr} 
\and
APC, AstroParticule et Cosmologie, Universit\'e Paris Diderot, CNRS, CEA, Observatoire de Paris, Sorbonne Paris , France
\label{inst2}
\and  
Institut d'Astrophysique de Paris, 98bis bd Arago, FR-75014 Paris,
  France \label{inst3} 
}
 
\abstract{
Cosmic rays are an essential ingredient in the evolution of
  the interstellar medium, as they dominate the ionisation of the
  dense molecular gas, where stars and planets form. However, since
  they are efficiently scattered by the galactic magnetic fields, many
  questions remain open, such as where exactly they are accelerated,
  what is their original energy spectrum, and how they propagate into
  molecular clouds.  In this work we present new observations and
  discuss in detail {a} method that allows us to measure the cosmic ray
  ionisation rate towards the molecular clouds close to the W28
  supernova remnant. To perform these measurements, we use CO, HCO$^+$, and DCO$^+$
  millimetre line observations and compare them with {the predictions of} radiative
  transfer and chemical models {away from thermodynamical equilibrium}. The CO observations allow
  us to constrain the density, temperature, and column density towards each
  observed position, while the DCO$^+$/HCO$^+$ abundance ratios
  provide us with constraints on the electron fraction and,
  consequently, on the cosmic ray ionisation rate.  Towards positions
  located close to the supernova remnant, we find cosmic ray
  ionisation rates much larger ($\gtrsim100$) than {those} in standard
  galactic clouds. Conversely, towards one position situated at a
  larger distance, we derive a standard cosmic ray ionisation rate.
  Overall, these observations support the hypothesis that the
  $\gamma$ {rays} observed in the region have a hadronic origin. In
  addition, based on CR diffusion estimates, we {find} that the
  ionisation of the gas is likely due to $0.1 -1$~GeV cosmic
  rays. Finally, these observations are also in agreement with the
  global picture of cosmic ray diffusion, in which the low-energy tail
  of the cosmic ray population diffuses at smaller distances than the
  high-energy counterpart. }

\keywords{molecular clouds - cosmic rays - SNR - ionisation - individual objects: W28}

\date{Received 2014/04/19 ; accepted 2014/06/05}

\maketitle



\section{Introduction}

\begin{figure*}[t]
  \centering
    \includegraphics[height=0.33\textheight]{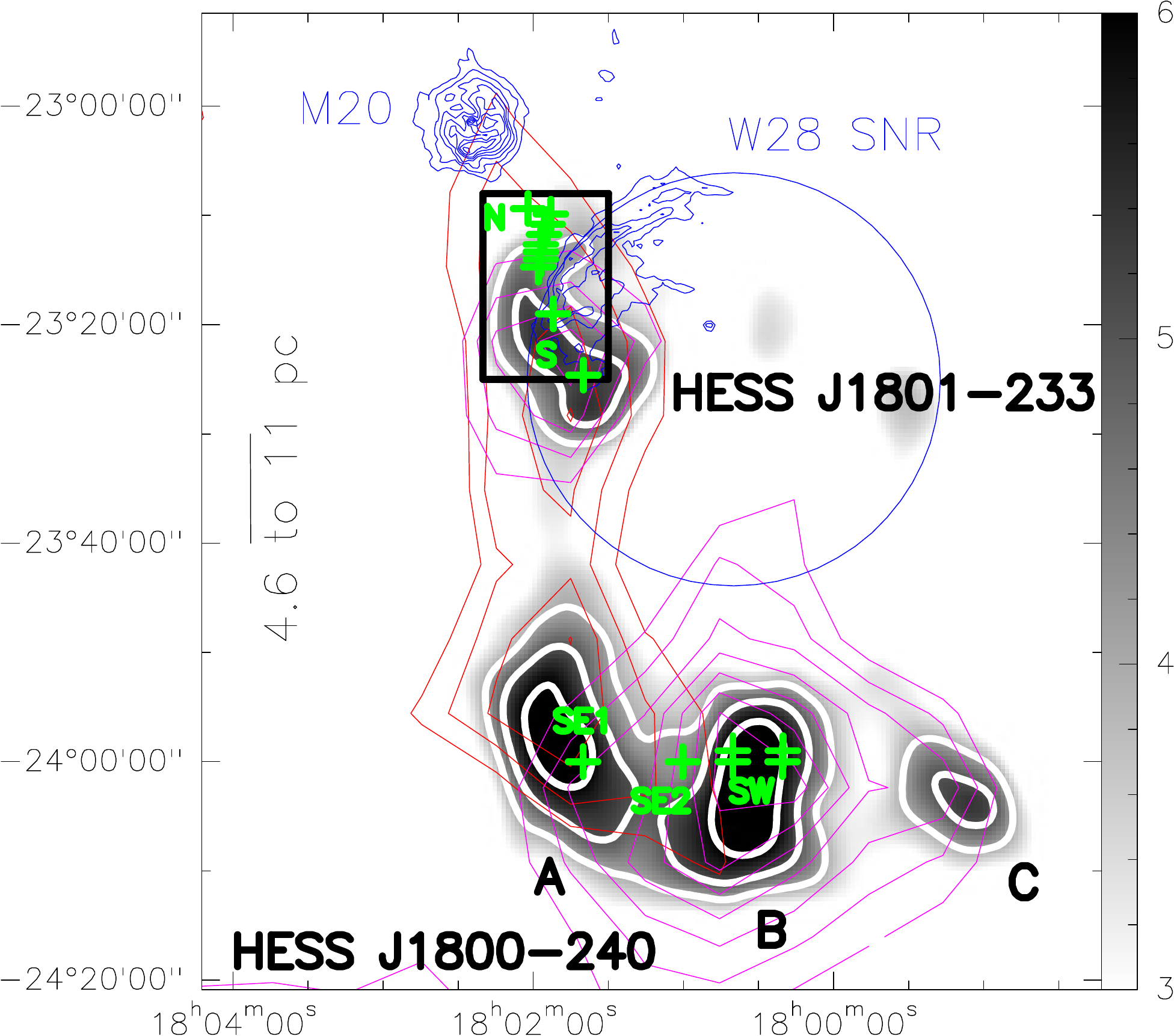}\hfill%
  \includegraphics[height=0.33\textheight]{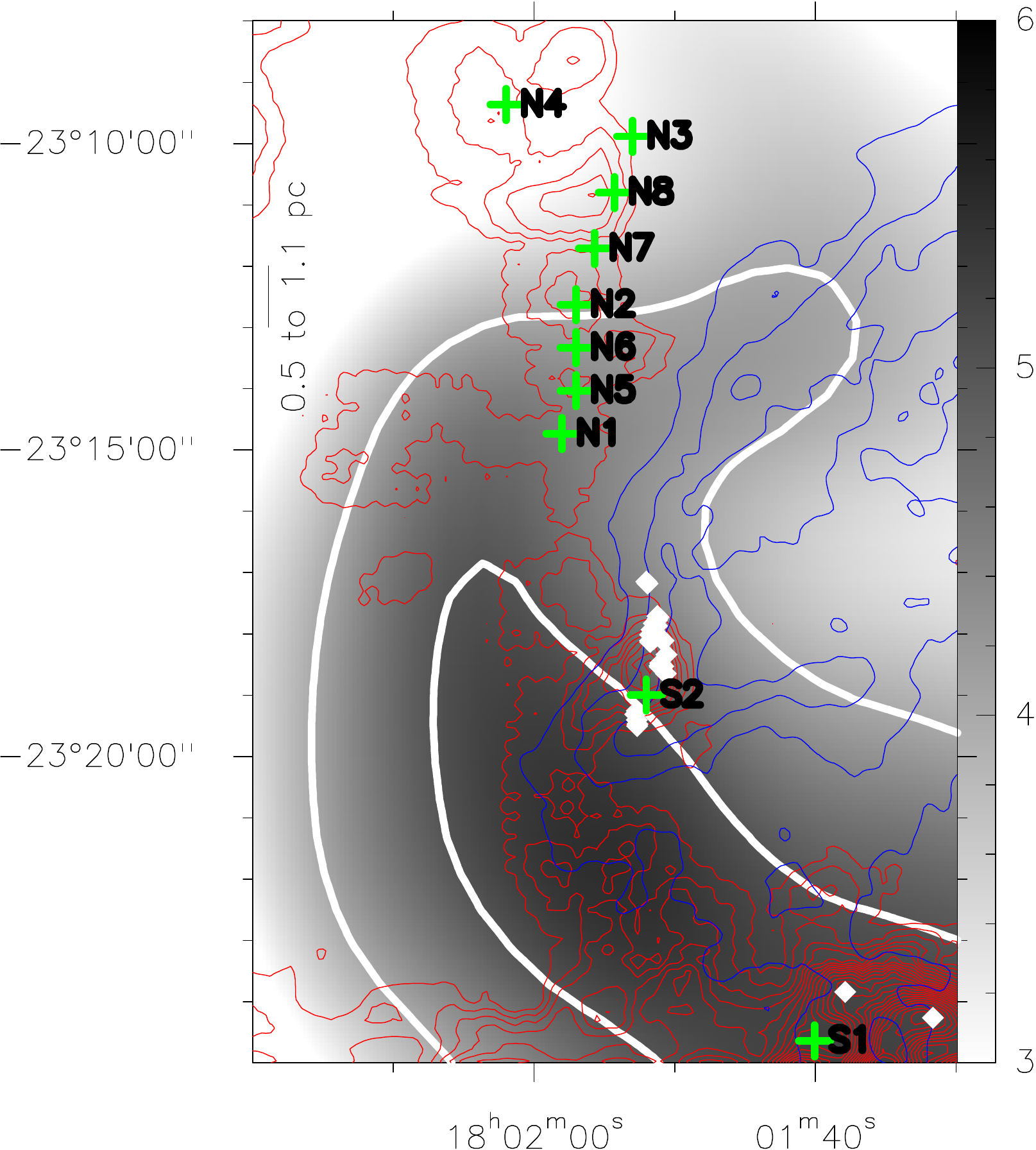}
  \caption{\textbf{(Left)} The W28 complex on large scales. 
  {Grayscale (in~$\sigma$) and thick contours} show TeV emission as seen by HESS  (levels are 4-6 $\sigma$). 
 {Red contours} show the CO(1-0) emission
    \citep{2001ApJ...547..792D} integrated over 15-25 \kms\ and {magenta
    contours} {trace} the emission integrated over 5-15 \kms\ (levels are 40-70~K~\kms\ by 5~K~\kms). Crosses show the positions observed {with} the IRAM 30m telescope and discussed in this paper. The blue contours show the 20~cm
    free-free emission in the M20 region \citep{Yusef-Zadeh2000}. The
    blue circle gives the approximate radio boundary of the SNR W28
    \citep{2006ApJ...639L..25B}.  \textbf{(Right)} The northern cloud
    in the W28 complex (zoom on the {black} box).  The {red contours} show the
    $\co (3-2)$ emission in K~\kms, integrated over 15-25~\kms\ {(levels are 15-130~K~\kms\ by 5~K~\kms)}
    \citep{Lefloch2008}. Diamonds show the locations of OH masers
    in the region \citep{Claussen1997}.}
  \label{fig:map}
\end{figure*}

Cosmic rays (CRs) are energetic charged particles that reach the Earth as an isotropic flux. They pervade the Galaxy and play a crucial role {in}
the evolution of the interstellar medium, because they dominate the
ionisation of  molecular clouds where the gas is shielded from the
UV radiation field.
The ionisation degree in the molecular gas is a fundamental
parameter {throughout} the star and planet forming process. First, ions
 couple the gas to the magnetic {field} and, therefore, they
regulate the gravitational collapse of the cloud. In addition, ions
sustain turbulence within protoplanetary discs and introduce non-ideal magnetohydrodynamics effects, which influence the accretion rate onto the protostar
\citep{1998RvMP...70....1B,2014arXiv1402.4133L}. Finally, the
CR induced ions initiate efficient chemical reactions in the cold
molecular clouds, which  eventually lead to the formation of complex
molecules, which enrich the gas even up to the first stages of planet
formation.

However, in order to fully understand the influence of CRs on the
{above} processes across the Galaxy, it is necessary to know where CRs are accelerated and how they
propagate through the gas.  Unfortunately, since CRs are scattered by
magnetic fields all along their {path} through the Galaxy, the production
  sites of CRs cannot be observed directly, and the diffusion of CRs
  also makes the evolution of the energy spectrum during their
  propagation {difficult} to determine observationally.

However, we can detect indirect signatures of the interaction of
  hadronic  CRs (essentially protons) {with matter}. Protons above {a
kinetic} energy threshold of $\approx280$~MeV produce $\pi^0$ pions when they
collide with {particles in the molecular cloud}. {Each pion then decays} into two \ensuremath{\gamma}-ray\
photons $(\pi^0 \rightarrow 2\gamma)$, each with a typical energy that thoseis $\sim
10\%$ that of the colliding proton. {Bright \ensuremath{\gamma}-ray\ sources thus indicate regions with a large density of protons with energies above 0.28 GeV. In these regions, the observed \ensuremath{\gamma}-ray\ photon spectrum can, in addition, be used to derive the spectrum of the parental CR particles, before the scattering and propagation within the Galaxy \citep{2013Sci...339..807A}.}

Supernova remnants (SNR) are thought to be the sources
of CRs. In this scenario, protons are accelerated in the expanding
shell of the SNR, following the diffusive shock acceleration process
\citep{1978MNRAS.182..147B}. Supporting this scenario, there is now
clear evidence that SNR are spatially associated with GeV to TeV
sources \citep{2013APh....43...71A}. Moreover, several SNR are close to
the molecular cloud that gave birth to the SN precursor. These
molecular clouds now act as reservoirs of target material for the freshly
accelerated protons, thus enhancing the production rate of \ensuremath{\gamma} rays.
Although it is compatible with \ensuremath{\gamma}-ray\ observations,
  the hadronic scenario is challenged by the leptonic scenario
  involving electron CRs. In this alternative scenario, the \ensuremath{\gamma}-ray\
emission can be explained mainly by inverse Compton scattering {of} the
cosmic microwave background
\citep[e.g. ][]{2009MNRAS.392..240M,2011ApJ...734...28A}.  Yet, this
scenario cannot explain the spatial correlation of TeV emission with
molecular clouds.  Moreover, recent observations of the IC443 and W44
SNR with the \textit{Fermi}-LAT telescope \citep{2013Sci...339..807A}
specifically support a hadronic origin of \ensuremath{\gamma} rays, consistent with the
so-called \textit{SNR paradigm} for the origin of primary CR
\cite[see e.g. ][ for a review]{Hillas2005}.

Cosmic ray protons with {kinetic} energy below the $\approx280$~MeV
  threshold of $\pi^0$ production cannot be traced by the emission of
\ensuremath{\gamma}-rays. Nevertheless, recent
  calculations suggest that the ionisation of UV-shielded gas is
  mostly due to keV-GeV protons \citep{padovani2009}. Accordingly,
  low-energy CR protons can be traced indirectly by measuring the
  ionisation fraction of the dense gas. It has thus been proposed
that an enhanced electron abundance in molecular clouds located in the
vicinity of SNR could be the smoking gun {for} the presence of freshly
accelerated CRs, with energies $\lesssim 1$~GeV.

This idea was put forward by \citet[][hereafter CC2011]{CC2011}, who
measured the ionisation fraction $\xe = n(\elec)/\nh$ in the W51C
molecular cloud, located in the vicinity of the W51 SNR. The detection
of TeV emission by both HESS and MAGIC telescopes close to the
molecular cloud is evidence of a physical interaction with the
SNR. This supports the idea of the pion-decay production of \ensuremath{\gamma} rays
with W51C acting as a \ensuremath{\gamma}-ray\ {emitter}. In CC2011, an enhanced
ionisation fraction was reported towards one position, W51C-E, which
required a CR ionisation rate two orders of magnitude larger than the
typical value of $1\times10^{-17}$~\s\ in molecular
clouds. This observational evidence strongly supports the
hadronic scenario of \ensuremath{\gamma}-ray\ production, at least {for} W51.

{Complementary studies of the CR ionisation rate in several diffuse clouds close to SNR have been carried out using different techniques, such as \htrp\ absorption \citep{2003Natur.422..500M}.
These studies also show an enhancement of a factor of 10-100 of the CR ionisation rate \citep{Indriolo:2010ul,2012ApJ...745...91I} with respect to the canonical value. However, the interpretation is not straightforward, as \citet{padovani2009} showed that the penetration into the cloud of high energy CRs results into an enhanced CR ionisation rate in low density molecular clouds even in absence of an increased CR flux.
}

{The combined} observations of two extreme energy ranges,
namely TeV and millimetre, seems a powerful method to characterise an
enhanced concentration of proton CRs. It
  also gives additional evidence {supporting} a physical interaction of the SNR shock with  molecular clouds.  From a theoretical point of view, it is
expected that the most energetic CR protons diffuse at larger
distances ahead of the SNR shock front, whilst the low-energy
tail of the distribution remains closer.  As
a consequence, one expects that any ionisation enhancement by low
energy CRs should be localised accordingly.  In CC2011, however, only
one location could be used to derive the ionisation fraction, and no
constraint could be given regarding the spatial distribution
  of the ionisation and therefore the diffusion properties of CRs.

The aim of this paper is to present measurements of the ionisation
fraction within the molecular clouds in the vicinity of the W28
SNR. The paper is organized as follows. In Section~2, the W28
association is presented, with particular emphasis on the physical
link between the SNR and the molecular clouds. In Section~3, the
millimetre observations are described. The derivation of the physical
conditions is presented in Section~4. The derivation of the ionisation
fraction and {the CR ionisation rates are}  described in Sections~5 and 6,
where we stress the strengths and limitations of the method. {The results}
are discussed in Section~7.

\section{The W28 association}
\label{sec:W28}
%

The W28 SNR has an age {greater} than $10^4$~yr, and is likely in the
Sedov or radiative phase \citep{Westerhout1958, Lozinskaya1974}. Its
distance is estimated between 1.6~kpc {and} 4~kpc, based on kinematic
determinations and H$\alpha$ observations \citep{Goudis1976,
  1981SvAL....7...17L}. The LSR velocity, based on H$\alpha$ and [NII]
observations, is estimated to be $18\pm5$~\kms\
\citep{Lozinskaya1974}. In the remainder of the text, all velocities
will refer to the local standard of rest (LSR) and projected distances will
be given for distances of both 1.6 and 4~kpc. {The boundary}
diameter of the SNR is 42~arcmin, corresponding to a linear radius of
9.6 to 24~pc (Fig.~\ref{fig:map}). 

The large-scale region towards W28
contains a variety of objects such as HII regions (e.g. M8, M20,
W28A2), molecular clouds, SNR, and new SNR candidates
\citep{2006ApJ...639L..25B}. Molecular gas, as seen in CO(1-0)
\citep{Wootten1981, 2001ApJ...547..792D}, coincides spatially with the
W28 SNR, suggesting a physical association, and supporting a view in
which the W28 SNR is interacting with its parental molecular
cloud. Probably related is the fact that ongoing massive star
formation has been observed in these molecular clouds, consistent with
the triggered star formation scenario \citep{1998ASPC..148..150E}. The
molecular gas located towards the north-east of the SNR boundary was
mapped at high spatial resolution, in the CO(3-2) rotational line, by
\cite{Lefloch2008} revealing a fragmented filamentary structure
elongated north-south (Fig.~\ref{fig:map}).

In \ensuremath{\gamma} rays, the high spatial resolution H.E.S.S. imaging array of
Cherenkov telescopes has revealed the presence of extended TeV
emission \citep{2008A&A...481..401A}, which splits into two components,
separated by 14-34~pc: HESS J1801-233, in the north, and HESS
J1800-240 in the south. The latter further splits into three
well-separated components (Fig.~\ref{fig:map}). In projection, the
entire TeV emission coincides with the molecular gas seen in CO(1-0)
which also appears to bridge the northern and southern TeV components.
The molecular gas coinciding with the northern TeV component J1801-233
shows  velocities predominantly from 15~\kms\ to 25~\kms, as does the
southern J1801-240~A component observed in CS(1-0) by
\cite{Nicholas2012}.  However, the southern components J1801-240~B and C
coincide with molecular emission characterised by somewhat lower
velocities, from 5~\kms\ to 15~\kms.  

Whether the TeV emission is physically
associated with the molecular clouds is of utmost importance, {for the question} of pion-decay production. However, there are several
indications, based on kinematic {information}, that this may well be
the case. 

First, when inspected in velocity space, the CO(1-0)
emission  covers the entire range from 5 to 25~\kms\ {continuously}
\citep{Fukui2012}. This indicates that the molecular emission, traced
either by CS or CO, is physically linked over the entire region, and
not only in projection. 

Second, {OH masers have been reported} towards the
northern component \citep{Claussen1997, Hewitt2008}, with velocities
ranging from 7.1~\kms\ to 15.2~\kms. Such OH masers are thought to trace the
interaction of the SNR shock with the molecular gas, an interpretation
which is consistent with the velocity range of the CO(1-0)
emission. 

Finally, velocity differences observed between the northern
and southern clouds are compatible with {the} differences up to
$\sim$6~\kms\ observed in the M20 map of CO(3-2), encompassing the
northern TeV component \citep{Lefloch2008}, indicating that velocity
shifts of several \kms\ are found which could be due to an interaction
with the SNR.  

Taken all together, these {data strongly} suggest a 3D
picture in which the SNR is interacting with surrounding molecular
clouds covering a wide and continuous velocity range, typically from 5~\kms\
to 25~\kms, and characterised by large variations along the line of
sight.

It is therefore most likely that the W28 association displays the
interaction of the SNR with molecular clouds, and thus {is} an
excellent target {in which} to study the ionisation by
CRs. 

\section{Observations and data reduction}
\label{sec:obs}


Observations were carried out over 40 hours in December, 2011, and
March, 2012, with the IRAM 30m~telescope.  We used the EMIR bands with
the Fast Fourier Transform Spectrometer as a backend in the
position-switching mode, using OFF positions about 1000" to the
east. {Amplitude} calibration was done typically every 15 minutes,
and pointing and focus were checked every 1 and 3 hours, respectively,
ensuring $\approx$2\arcsec\ pointing accuracy. All spectra were
reduced using the CLASS package of the
GILDAS\footnote{\url{http://www.iram.fr/IRAMFR/GILDAS/}} software
\citep{2005sf2a.conf..721P}. Residual bandpass effects were subtracted
using low-order ($\leq 3$) polynomials. The weather was good and
$T_\text{sys}$ values {were} lower than 220~K. The observed molecular
transitions used in the present work are listed in
Table~\ref{tab:species}, along with the associated system temperature
and sensitivity ranges $T_\text{sys}$ and $\sigma_{rms}$ obtained
during the successive runs of observations. All spectra are presented
{on} the main-beam temperature scale, $T_{\rm mb} = (F_{\rm eff}/B_{\rm eff})T_{\rm ant}^*$, {with $F_{\rm eff}$ and $B_{\rm eff}$ the forward and main-beam efficiencies of the telescope, respectively.}

We observed towards 16 positions, {of} which 10 are located in the
northern cloud and 6 in the southern cloud. The coordinates of these
positions are listed in Table~\ref{tab:positions}. The two lowest
rotational transitions of \thco\ and \cdo\ are used to derive the
physical conditions in the cloud, while \htrcop(1-0) and \dcop(2-1)
are used to derive the cosmic ray ionisation rate.


\begin{table*}
  \caption{\label{tab:species} Molecular transitions observed with
    the IRAM 30m telescope.}
  {\centering
  \begin{tabular}{lcrccccc}
    \toprule
    Species & Line & Frequency &$F_{\rm eff}$&$B_{\rm eff}$ & HPBW & $T_{\rm sys}$&
    $\sigma_{rms}$\\
    &&[GHz]&&&[arcsec]&[K]&[mK]\\
    \midrule
    \htrcop&(1-0)&86.754&{0.95}&0.81&29&$100-130$&$6-12$\\
    \cdo&(1-0)&109.782&{0.95}&0.79&22&$140-200$ &$20-50$\\
    \thco&(1-0)&110.201&{0.95}&0.79&22&$140-200 $&$20-50$\\
    \cseo&(1-0)&112.359&{0.95}&0.79&22&$ 140-200$&$20-50$\\
    \dcop&(2-1)&144.077&{0.92}&0.74&16&$100-200 $&$8-20$\\
    \cdo&(2-1)&219.560&{0.94}&0.61&11&$160-220 $&$20-80$\\
    \thco&(2-1)& 220.399&{0.94}&0.61&11& $ 160-220$&$20-80$\\
    \cseo&(2-1)&224.714&{0.94}&0.61&11&$160-220 $&$20-80$\\
    \bottomrule
  \end{tabular}
  \\ }
  \begin{mytablecaption}
    $T_\text{sys}$ indicates the range of system temperatures during
    the observing run, and the corresponding sensitivity
    fluctuations. The adopted values of the telescope parameters
    follow from the IRAM observatory recommendations: {$F_{\rm eff}$ and $B_{\rm eff}$ are the forward and main-beam efficiencies of the telescope, respectively}; HPBW is the half-power beam
    width.
  \end{mytablecaption}
\end{table*}

\begin{table}[t]
  \caption{J2000 coordinates of the 16 observed positions.  \label{tab:positions}}
  \centering
    \begin{tabular}{lrrl}
      \toprule
      Source & $\alpha_{2000}$&$\delta_{2000}$\\
             & $(h m s)$ & $(\degr\,\arcmin\,\arcsec)$ \\
      \midrule
      J1801-N1 & 18 01 58.0 & -23 14 44 \\
      J1801-N2$\dagger$ & 18 01 57.0 & -23 12 38 \\
      J1801-N3$\dagger$ & 18 01 53.0 & -23 09 53 \\
      J1801-N4$\dagger$ & 18 02 02.0 & -23 09 22   \\
      J1801-N5 & 18 01 57.0 & -23 14 02 \\
      J1801-N6 & 18 01 57.0 & -23 13 20 \\
      J1801-N7 & 18 01 55.7 & -23 11 43 \\
      J1801-N8 & 18 01 54.3 & -23 10 48 \\
      J1801-S1 & 18 01 40.0 & -23 24 38\\
      J1801-S2 & 18 01 52.0 & -23 19 00\\
      J1801-SE1 & 18 01 40.0 & -24 00 00\\
      J1801-SE2 & 18 01 00.0 & -24 00 00\\
      J1801-SW1 & 18 00 40.0 & -24 00 00\\
      J1801-SW2 & 18 00 20.0 & -24 00 00\\
      J1801-SW3 & 18 00 40.0 & -23 59 00\\
      J1801-SW4 & 18 00 20.0 & -23 59 00\\
      \bottomrule
    \end{tabular}
  \begin{list}{}{}
  \item $\dagger$ N2, N3, and N4 correspond to TC5, TC7, and TC6,
    respectively, as referred to in \cite{Lefloch2008}. N2 also
    coincides with a high-mass protostar.
  \end{list}
\end{table}

%
\section{Results}
\label{sec:results}
%

\subsection{Observed spectra}

\begin{figure*}
  \centering
  \includegraphics[width=0.95\hsize]{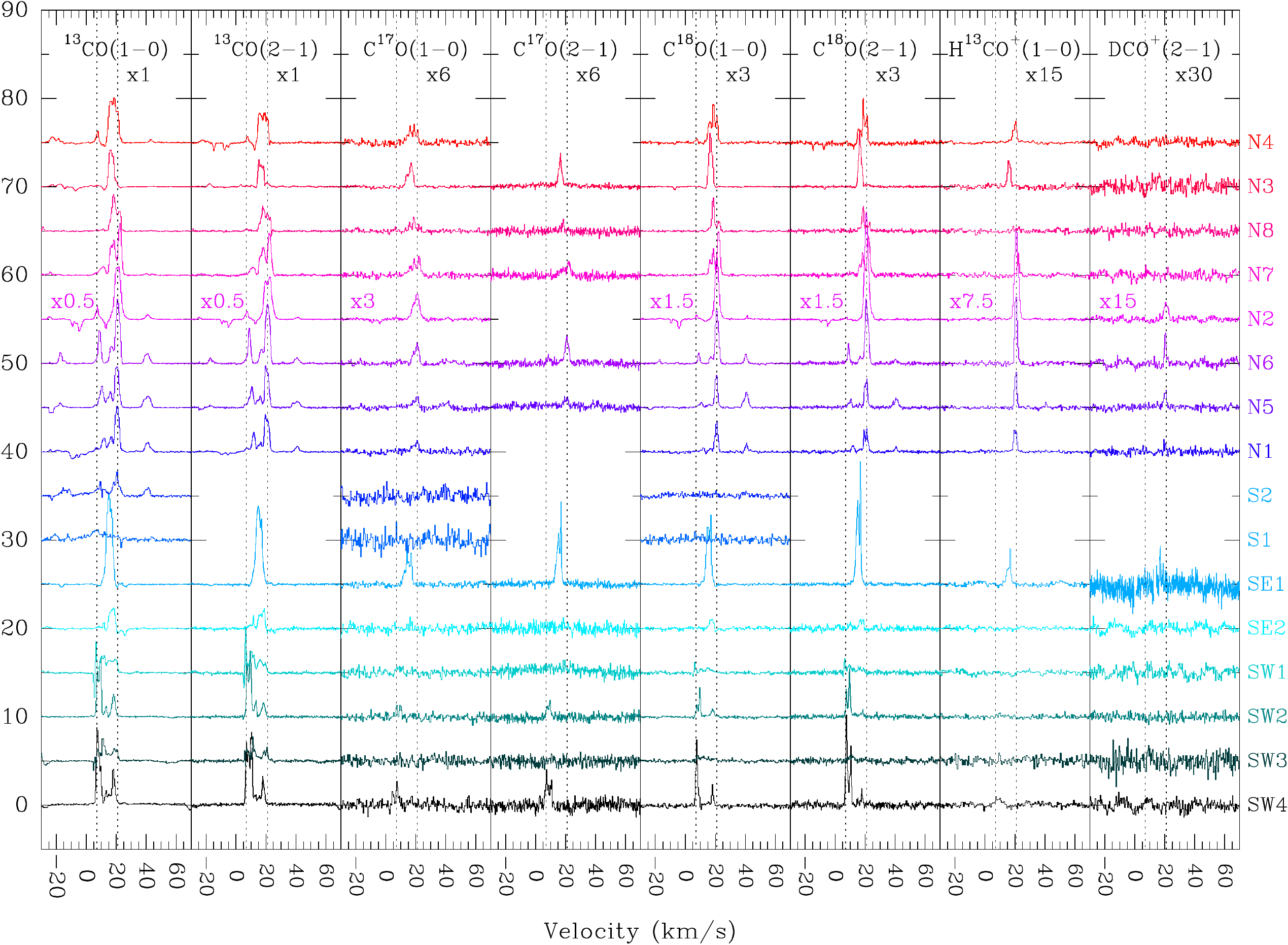}
  \caption{Observations of millimetre 
    emission lines towards all positions. The intensities are in units
    of main-beam temperature~(K). For readibility, a multiplicative
    factor was applied to the spectra and is given under each
    transition line label. This factor was decreased for position N2 (protostar)
    where the signal is very strong. Vertical dashed lines indicate the two
    extreme velocity components of the complex at 7 and 21~\kms.}
  \label{fig:spectra}
\end{figure*}

The resulting spectra towards all positions are shown in
Fig.~\ref{fig:spectra}. The \thco\ and \cdo\ spectra show multiple
velocity components, most likely associated with several clouds along
the line of sight. In some instances, negative features are apparent,
which are due to emission from the reference position at different
velocities. Isotopologues \thco\ and \cdo\  are detected towards 12 of
the 16 positions, and the spectra show clearly two main components.
The velocity of the dominant component varies between the northern
($\lesssim 21$~\kms) and southern ($\gtrsim 7$\kms) cloud, as
presented in \textsection~\ref{sec:W28}.
At most positions, the rarer \ce{C^{17}O} isotopologue is also
detected, although the hyperfine structure of the (1-0) transition is
not always well resolved. The \htrcop(1-0) and \dcop(2-1) emission
lines have only one velocity component, at 21~\kms.  The \cdo(1-0) and
(2-1) lines show a clear distinction between the northern positions,
where the line emission is the strongest, and southern
positions. The compound \htrcop\ is detected towards all northern positions but
N8. However, \dcop\ is detected only towards three positions, N2, N5,
and N6. Southern positions display much weaker {emission} and SE1 is
detected in both \htrcop\ and \dcop.

The analysis was essentially driven by the \htrcop\ and \dcop\ lines
which are the main {focus} of the present work. The {velocity of the} dominant CO transition always corresponds to the  velocity of the \htrcop\ line when detected.
When more velocity components are detected in CO, we limited the Gaussian fits to the first two dominant components.
Upper limits are given at the $1\sigma$ level for
$T_\text{peak}$, and for the integrated intensity $W$ by assuming $\Delta v=3$~\kms. The results from the Gaussian fits are summarised in
Table~\ref{tab:fits} (CO isotopologues) and Table~\ref{tab:hcop} (\htrcop, \dcop).

In the following, we describe the two-step analysis of the measured line
intensities. First, the physical conditions are derived
according to the \thco\ and \cdo\ lines. Second, the
\ce{HCO+}/\ce{DCO+} abundance ratio is compared with model predictions
computed using the derived physical conditions.

\begin{table*}[t]
  \caption{\label{tab:fits} Results from the Gaussian fits of the
    emission lines of  \thco\ and \cdo\ towards the 12 positions where they are detected.}
  {\centering
    \begin{tabular}{lrrrrrrrrrrrrrrrr}
      \toprule 
      Pos.& \multicolumn{4}{c}{\thco$(1-0)$} & \multicolumn{4}{c}{\thco$(2-1)$} &
      \multicolumn{4}{c}{\cdo$(1-0)$} & \multicolumn{4}{c}{\cdo$(2-1)$}\\   
      &$v_0$&$W$&$T_\text{peak}$&$\Delta v$&$v_0$&$W$&$T_\text{peak}$
      &$\Delta v$&$v_0$&$W$&$T_\text{peak}$&$\Delta v$&$v_0$&$W$&$T_\text{peak}$&$\Delta v$\\
      \midrule
            \multicolumn{7}{l}{First dominant velocity component}\\
      N1 &   20.7 &  17.4 &  5.3 & 3.1  &   20.5 &  15.1 & 4.1 & 3.5  &   20.9 & 2.7 & 1.0 & 2.4  &  21.1 & 1.5 & 0.8 & 1.9  \\
      N5 &   20.6 &  15.7 &  5.0 & 3.0  &   20.5 &  15.0 & 4.7 & 3.0  &   20.8 & 2.8 & 1.3 & 2.1  &  20.7 & 2.6 & 1.0 & 2.4  \\
      N6 &   21.2 &  23.9 &  7.4 & 3.0  &   21.1 &  24.2 & 7.0 & 3.3  &   21.2 & 5.5 & 2.0 & 2.5  &  21.1 & 6.2 & 2.3 & 2.5  \\
      N2 &   21.1 &  59.6 & 12.6 & 4.5  &   21.5 &  57.4 & 9.6 & 5.6  &   21.1 &   14.7 & 4.5 & 3.0  &  21.0 &   25.0 & 7.9 & 3.0  \\
      N7 &   22.5 &  15.3 &  5.8 & 2.5  &   22.4 &  13.9 & 4.8 & 2.7  &   22.3 & 3.6 & 1.7 & 1.9  &  22.3 & 3.4 & 1.5 & 2.1  \\
      N8 &   18.4 &  15.4 &  4.0 & 3.6  &   18.2 &   9.7 & 2.5 & 3.6  &   18.6 & 2.4 & 1.3 & 1.7  &  18.6 & 1.3 & 0.9 & 1.4  \\
      N3 &   16.8 &  17.3 &  4.4 & 3.7  &   16.6 &  11.6 & 2.8 & 3.9  &   16.8 & 5.3 & 1.9 & 2.6  &  16.7 & 4.6 & 1.9 & 2.3  \\
      N4 &   18.9 &  15.2 &  4.9 & 2.9  &   19.0 &  11.3 & 3.3 & 3.2  &   18.9 & 2.2 & 1.5 & 1.4  &  18.8 & 1.4 & 1.6 & 0.8  \\
      SE1 &   15.0 &   34.3 & 9.9 & 3.3  &   15.1 &   39.7 & 8.7 & 4.3  &   15.3 & 6.7 & 2.1 & 3.0  &   15.0 & 7.5 & 3.0 & 2.4  \\
      SE2 &   16.3 & 5.8 & 2.0 & 2.7  &   16.3 & 5.6 & 1.6 & 3.2  &   16.9 & 0.8 & 0.3 & 2.5  &   16.9 & 0.6 & 0.3 & 1.9  \\
      SW2 & 9.6 &   10.5 & 7.0 & 1.4  & 9.5 &   13.2 & 7.4 & 1.7  & 9.7 & 1.3 & 1.1 & 1.1  & 9.7 & 2.2 & 1.8 & 1.2  \\
      SW4 & 8.9 &   14.7 & 4.7 & 3.0  & 7.8 &   18.6 & 6.4 & 2.7  & 7.5 & 3.4 & 2.4 & 1.4  & 7.6 & 4.6 & 3.1 & 1.4  \\
      \midrule
      \multicolumn{7}{l}{Second dominant velocity component}\\
      N1 &   11.7 &  5.6 & 1.5 & 3.6  &   11.8 &   5.7 & 2.2 & 2.5   &   11.9 & 0.3 & 0.2 & 1.8  &   11.9 & 0.5 & 0.2 & 2.1  \\
      N5 &   10.4 &  8.0 & 2.3 & 3.3  &   10.4 &   7.3 & 2.1 & 3.3  &   10.4 & 0.6 & 0.2 & 3.2  &  10.1 & 0.8 & 0.3 & 3.0  \\
      N6 & 9.0 &  8.3 & 3.7 & 2.1  & 9.0 &   8.1 & 3.9 & 2.0  & 9.0 & 0.6 & 0.4 & 1.5  &  9.0 & 1.0 & 0.7 & 1.4  \\
      N7 &   18.6 &  10.0 & 3.9 & 2.4  &   18.3 &   9.1 & 3.2 & 2.7  &   18.7 & 1.8 & 1.0 & 1.8  &  18.6 & 1.6 & 0.8 & 1.9  \\
      N8 &   22.3 &  4.5 & 2.2 & 1.9  &   21.3 &   3.0 & 1.6 & 1.8  &   22.3 & 0.9 & 0.3 & 2.3  &  22.1 & 0.8 & 0.2 & 3.3  \\
      N4 &   16.0 &  12.4 & 4.5 & 2.6  &   15.8 &   8.3 & 3.4 & 2.3  &   16.3 & 2.3 & 0.8 & 2.6  &  16.1 & 1.5 & 0.5 & 2.5  \\
      SE1 &   17.3 &   10.4 & 4.6 & 2.1  &   17.2 & 1.9 & 2.6 & 0.7  &   17.0 & 1.6 & 1.9 & 0.8  &   17.0 & 3.0 & 4.3 & 0.7  \\
      SE2 &   18.6 & 4.3 & 2.2 & 1.9  &   18.8 & 3.0 & 1.9 & 1.5  &   18.5 & 0.3 & 0.2 & 1.5  &   18.8 & 0.4 & 0.3 & 1.1  \\
      SW2 & 7.6 & 9.1 & 5.3 & 1.6  & 7.5 & 8.0 & 6.2 & 1.2  & 7.6 & 0.6 & 0.4 & 1.3  & 7.5 & 1.0 & 1.0 & 0.9  \\
      SW4 & 7.3 &   11.8 & 6.9 & 1.6  &   10.4 &   12.7 & 7.9 & 1.5  &   -&-&-&-  &   10.5 & 2.5 & 1.9 & 1.2  \\
      \bottomrule
    \end{tabular}
    \\ }
  \begin{mytablecaption}
    The fit parameters are: the centre line velocity $v_0$ (in \kms,
    in the local standard of rest), the integrated intensity $W$ (in
    K~\kms), the peak temperature $T_\text{peak}$ (in K), and the full
    width at half maximum $\Delta v$ (in \kms). In case of
    non-detections, upper limits are given at the 1~$\sigma$ level.
    Uncertainties are dominated by calibration ($\sim 20$~\%).  {Integrated intensities and peak temperatures are given on the main-beam temperature scale}.
  \end{mytablecaption}
\end{table*}

%
\subsection{Determination of  physical conditions}
\label{sec:physcond}
%
\begin{table*}[t]
  \caption{\label{tab:hcop} Results from the Gaussian fits of the
    emission lines of  \htrcop\ and \dcop \ towards the nine positions where \htrcop\ is detected.}
  {\centering
    \begin{tabular}{l rrrr rrrr}
    \toprule 
    Pos. & \multicolumn{4}{c}{\htrcop$(1-0)$} & \multicolumn{4}{c}{\dcop$(2-1)$}\\
    &$v_0$& $W$ & $T_\text{peak}$ & $\Delta v$
    &$v_0$& $W$ & $T_\text{peak}$ & $\Delta v$
    \\
    \midrule
    N1 & 20.3 & 0.44 & 0.17 &  2.5  &  -  &  $<$0.01 &  $<$0.01 &  3.0  \\
    N5 & 20.6 & 0.53 & 0.27 &  1.8  & 20.4 & 0.15 & 0.06 &  2.3  \\
    N6 & 20.8 & 1.11 & 0.53 &  2.0  & 20.5 & 0.16 & 0.12 &  1.2  \\
    N2 & 21.0 & 4.39 & 1.38 &  3.0  & 21.0 & 0.41 & 0.12 &  3.1  \\
    N7 & 22.3 & 0.23 & 0.16 &  1.3  &  -  &  $<$0.01 &  $<$0.01 &  3.0  \\
    N3 & 16.2 & 0.49 & 0.20 &  2.3  &  -  &  $<$0.02 &  $<$0.02 &  3.0  \\
    N4 & 20.2 & 0.47 & 0.14 &  3.0  &  -  &  $<$0.01 &  $<$0.01 &  3.0  \\
    SE1 & 17.0 & 0.15 & 0.23 &  0.6  &  17.0  &  0.13  &  0.16  &  0.76 \\
    SW4 &  9.3 & 0.25 & 0.05 &  5.2  &  -  &  $<$0.02 &  $<$0.02 &  3.0  \\
    \bottomrule
  \end{tabular}
  \\ }
\begin{mytablecaption}
  The fit parameters are the same as Table~\ref{tab:fits}.  In case of
  non-detections, upper limits are given at the 1~$\sigma$ level.
  Uncertainties are dominated by calibration ($\sim 20$~\%).
    {Integrated intensities and peak temperatures are given on the main-beam temperature scale}
\end{mytablecaption}
\end{table*}

The physical conditions prevailing at the various locations were
determined based on the \thco\ (1-0) and (2-1) lines and the \cdo\ (1-0) and (2-1) lines. To {accomplish this},
 we performed non-LTE (local thermal equilibrium) calculations of the rotational level
populations under the large velocity gradient (LVG) approximation
\citep{CC2003}. The \hh\ density, gas kinetic temperature \tkin, and
total column density of each species covered a large parameter
space. For each set of input parameters, the expected line intensities
and integrated intensities were computed adopting the linewidth
$\Delta v$ derived from the Gaussian fits (Table~\ref{tab:fits}),
and taking into account beam dilution effects by varying the size of
the emitting regions.  A simple $\chi^2$ minimization was then used to
constrain the physical conditions that best reproduce the observed
intensities. The LVG analysis was performed towards positions
where all four lines were detected. We considered collisions of \thco\
and \cdo\ with both para and ortho \hh, assuming an ortho-to-para ratio in Boltzmann equilibrium.
Hence, in the temperature range considered here, \hh\ is  mainly in its para configuration.
We used the collisional cross sections from \cite{2010ApJ...718.1062Y}.

In this process, the relative abundances of \cdo\ to \thco\
need to be known, and we have assumed that the molecular isotopic
ratios are equal to the elemental isotopic ratios, namely that
\begin{equation}
  \frac{[\cdo]}{[\thco]} = \frac{[\cdo]}{[\docseo]} \times 
  \frac{[\docseo]}{[\thco]} 
  \approx \left(\frac{\huo}{\seo}\right) \times \left(\frac{\doc}{\trc}\right).
\end{equation}
It is known that isotopic ratios may vary with the position within the
Milky Way, resulting from the gradual depletion of \doc\ and
enrichment of \trc\ with the cycling of gas through
stars. \citep[e.g. ][]{1994ARA&A..32..191W, 1982ApJ...262..590F}.  In
addition, local variations are also possible, resulting from the
competition of chemical fractionation and selective photodissociation
\citep{1988ApJ...334..771V,2003ApJ...591..986F}. The dependence of the
\doc/\trc\ isotopic ratio on galactocentric distance was studied by
\cite{2005ApJ...634.1126M}.  Applying their results to W28, which
is 4-6~kpc from the Galactic center, one  gets
$\doc/\trc=50\pm7$. In practice, in our non-LTE analysis, we varied
the $\trc/\doc$ ratio. The best $\chi^2$ values were obtained using an
isotopic ratio of 50, consistent with the above expectation, 
which we therefore adopted in what follows. Regarding the $\seo/\huo$
isotopic ratio, we adopted the  value of 500
representative of the solar neighborhood.

The results of the LVG analysis are summarised in
Table~\ref{tab:results}. At each position, the size of the emitting
regions was found to be larger than the beam size.  Position N2, which
is considered a protostar in \citet{Lefloch2008}, shows a peculiar behaviour with a
visual extinction at least 5~times higher than for the other
positions.
Overall, the density and kinetic temperature we
derived are typical of dense molecular clouds with visual extinctions
larger than 10 magnitudes. We also note that the \hh\ densities we derived in the northern cloud are
consistent with values published by \citet{Lefloch2008}.

\begin{table*}[t]
  \caption{\label{tab:results}
    Physical conditions and  cosmic ray ionisation rates.
   }
  {\centering
  \resizebox{\hsize}{!}{%
    \begin{tabular}{lcrrrrrrrr}
      \toprule
      Pos.&$\Delta v$&$n_{\hh}$&$T_{kin}$ & $N (\cdo)$&$A_V$& $N (\htrcop)$& $N (\dcop)$ &$\rd = \frac{[\dcop]}{[\hcop]}$&$\zeta$\\
     &[\kms]&[$10^3$\ \cmmm]&[K]&[$10^{15}$\ \cmm]&[mag]&[$10^{12}$\ \cmm]&[$10^{12}$\ \cmm]&&[$10^{-17}$ \s]\\
      \midrule
      N1&  3.5&  $0.6\ \{0.2-1\}$&$15\pm5$&$4\ \{2-6\}$  &$21\ \{11-32\}$ & $0.8 - 1.3$&$ < 0.22$&$ < 0.005$&$>$  13 \\
      N5&  3.0&$4\ \{2-5\}$&$10\pm2$&$3\ \{2-8\}$  &$16\ \{11-32\}$ & $1.1 - 1.4$&$0.89 -1.30$&$0.014 -  0.020$& 130 - 330 \\
      N6&  3.0&$4\ \{2-6\}$&$13\pm3$&$6\ \{4-20\}$ &$32\ \{21-105\}$& $1.8 - 2.5$&$0.79 -1.30$&$0.008 -  0.012$& 130 - 400 \\
      N2$^\dagger$&  5.0&  $>2$&$16\pm2$&  $20\ \{15-30\}$&  $105\ \{79-158\}$& $5.6 - 8.9$&$1.10 -2.00$&$0.003 -  0.006$&  - \\
      N7&  2.5&$2\ \{2-5\}$&$10\pm2$&$4\ \{3-10\}$ &$21\ \{16-53\}$ & $0.6 - 0.9$&$ < 0.25$&$ < 0.007$&$>$ 130 \\
      N8&  3.5& $1\ \{0.6-2\}$&$ 8\pm1$&$3\ \{2-4\}$  &$16\ \{11-21\}$ & $ < 0.2$&$ < 0.35$& -  &  -  \\
      N3&  3.5&  $6\ \{4-10\}$&$ 8\pm1$&$6\ \{5-7\}$  &$32\ \{26-37\}$ & $1.0 - 1.4$&$ < 0.35$&$ < 0.006$&$>$ 260 \\
      N4&  3.0& $2\ \{0.6-4\}$&$12\pm3$&$2\ \{2-3\}$  & $11\ \{5-16\}$ & $1.0 - 1.4$&$ < 0.35$&$ < 0.006$&$>$  40 \\
      SE1&  4.0&$2\ \{1-5\}$&$19\pm5$&  $6\ \{5-20\}$  &$32\ \{26-105\}$& $0.4 - 0.56$&0.79 - 1.0  &  0.032 - 0.05 &  0.2 - 20  \\
      SE2&  3.0&  $4\ \{2-10\}$&$ 8\pm2$&$0.9\ \{0.4-20\}$& $5\ \{2-105\}$ & $ < 0.2$&$ < 0.28$&  - &  -  \\
      SW2&  1.5&$2\ \{1-4\}$&$20\pm4$& $4\ \{3-10\}$&$21\ \{16-53\}$ & $ < 0.1$&$ < 0.22$&  - &  -  \\
      SW4$^\dagger$&  1.5&  $6\ \{4-10\}$&$16\pm2$&$1.5\ \{1-3\}$&  $5\ \{5-16\}$ & $0.5 - 0.8$&$ < 0.25$&$ < 0.009$&  -  \\
      \bottomrule
    \end{tabular}
  }%
  \\ }
\begin{mytablecaption}
  $n_\hh$ is the molecular hydrogen density (\cmmm), \tkin\ the gas
     kinetic temperature, $N(\cdo)$ the total column density of
     \cdo. $A_V$ is the visual extinction assuming $[\cdo] = A_V \times 1.9\ 10^{14}$ \cmm\ 
        \citep{1982ApJ...262..590F,2013ARA&A..51..207B}.  We 
     assumed isotopic ratio values $\huo/\seo=500$ and
     $\trc/\doc=50$ (see text). Values in brackets indicate the range of
     values satisfying $\chi^2_{\nu}<1$. 
     Uncertainties on $n_\hh$ and \tkin\
     are at the 70\% confidence level, and are propagated in the
     abundance ratios and upper limits. Lower limits of $\zeta$ were
     deduced from chemical modelling (see section \textsection
     \ref{sec:results_CRI}). \\ $^\dagger$ N2 and SW4 are probably  ionised by a source other than CRs (see text).
  \end{mytablecaption}
 \end{table*}

\subsection{The \dcop/\hcop\ abundance ratio}
The species \dcop\ was detected towards four positions, for which it was possible to derive {values of the $\dcop/\hcop$ abundance ratio}. For all other positions where \htrcop\ was detected,  upper limits  at the $1 \sigma$ level on the abundance ratio we derived.
We determined the column densities of \htrcop\ and \dcop\ from the same non-LTE LVG calculation, using the collisional cross sections from \cite{1999MNRAS.305..651F} and the physical conditions derived from the CO observations (Table~\ref{tab:results}). We derived the observed $\dcop/\hcop$ abundance ratio for each  set of physical conditions $(n_\hh,T)$, assuming $\doc/\trc=50$ (see \S~\ref{sec:physcond}). The uncertainty in {the abundance ratio} is dominated by the uncertainties in the physical conditions.  Results are listed in Table~\ref{tab:results} and will be used in the next section to constrain the CR ionisation rate.

%
\section{Methods to measure the CR ionisation rate $\zeta$}
\label{sec:CRI}
%


\subsection{Analytical method}\label{sec:analytical-method}

In their seminal paper, \citet[hereafter G77]{Guelin1977}  suggested
that the abundance ratio of \dcop\ to \hcop, which we denote $\rd =
\dcop/\hcop$, can be used to measure the ionisation fraction in
molecular clouds, $\xe = n(\elec)/\nh$. Subsequently,
\citet[hereafter C98]{Caselli1998}  proposed using the
$R_H=\hcop/\co$ abundance ratio in combination with \rd\ to derive
both \xe\ and $\zeta$ in dark clouds. The basic idea is
that  \dcop\ and \hcop\ are chemically linked, and
depend on a reduced number of chemical reactions where the CR
ionisation rate plays a crucial role, through the ionisation of \hh\
into \ce{H2+} \citep{1973ApJ...185..505H}, leading to the formation of the
pivotal \ce{H3+} ion. The fast ion-neutral reaction of \ce{H3+} with {HD}
produces the deuterated ion \ce{H2D+} which then initiates the
formation of several deuterated species, {including} \dcop. In a
similar fashion, \ce{HCO+} is formed by the reaction of \ce{H3+}
 with CO. This forms the basis of the method of C98 which uses
CO, \dcop, and \hcop\ to derive \xe\ and $\zeta$. The full set of
reactions used in the C98 analysis is given in
Table~\ref{tab:reactions} with updated reaction rates.

The steady-state abundance ratios $R_H=\hcop/\co$ and $\rd =
\dcop/\hcop$ can be analytically derived from this network,
provided that \hcop\ and \dcop\ are {predominantly} formed and destroyed
by reactions 1-4 and 7-9, respectively. One then finds that
\begin{equation}
  \label{eq:RH}
  R_H = \frac{[\hcop]}{[\co]} = \frac{k_H}{\beta'} \frac{x(\htrp)}{\xe}
  \approx \frac{k_H}{(2 \beta \xe + \delta) \beta'}\frac{\zeta/n_\h}{\xe}\ ,
\end{equation}
and
\begin{equation}
  \label{eq:RD2}
  \rd = \frac{[\dcop]}{[\hcop]} 
  \approx \frac{1}{3}\frac{x(\hhdp)}{x(\htrp)}
  \approx \frac{1}{3} \frac{k_f x(\hd)}{k_e \xe+\delta+k_f^{-1}/2} \ ,
\end{equation}
where $n(\ce{X})$ is the number density of species \ce{X} and $x(\ce{X}) =
n(\ce{X})/n_{\h}$ its fractional abundance. In the following, we will
assume that the gas is fully molecular such that $n_\h = 2n(\hh)$. The
coefficients $\beta$, $\beta'$, and $k$ are the reaction rates listed
in Table~\ref{tab:reactions}. Finally, $\delta \sim \delta_{\htrp} \sim \delta_{\hhdp}$ is the total
destruction rate of \htrp\ or \hhdp\ by neutrals. Provided that $R_H$, \rd,
\nh, and the kinetic temperature are known, \xe\ and $\zeta$ can then
be derived as
\begin{eqnarray}
  \xe &=& \left(\frac{k_f x(\hd)}{3 \rd}-\delta -
    \frac{k_f}{2} \text{e}^{-\Delta E/T} \right) \frac{1}{k_e}\label{eq:xe}\ ,\\
  \zeta/n_\h &=& \frac{\beta'}{k_H} \left(2 \beta \xe + \delta \right) R_H  \xe\ ,
  \label{eq:zeta}
\end{eqnarray}
where $\Delta E=220$~K, such that at sufficiently low temperatures
the last term in brackets in Eq.~\ref{eq:xe} becomes
negligible. Equation \ref{eq:xe} demonstrates that \xe\ only depends
on the abundance ratio \rd\ and the gas kinetic temperature, as
originally proposed by G77.

\begin{figure} 
  \centering
  \includegraphics[width=\hsize]{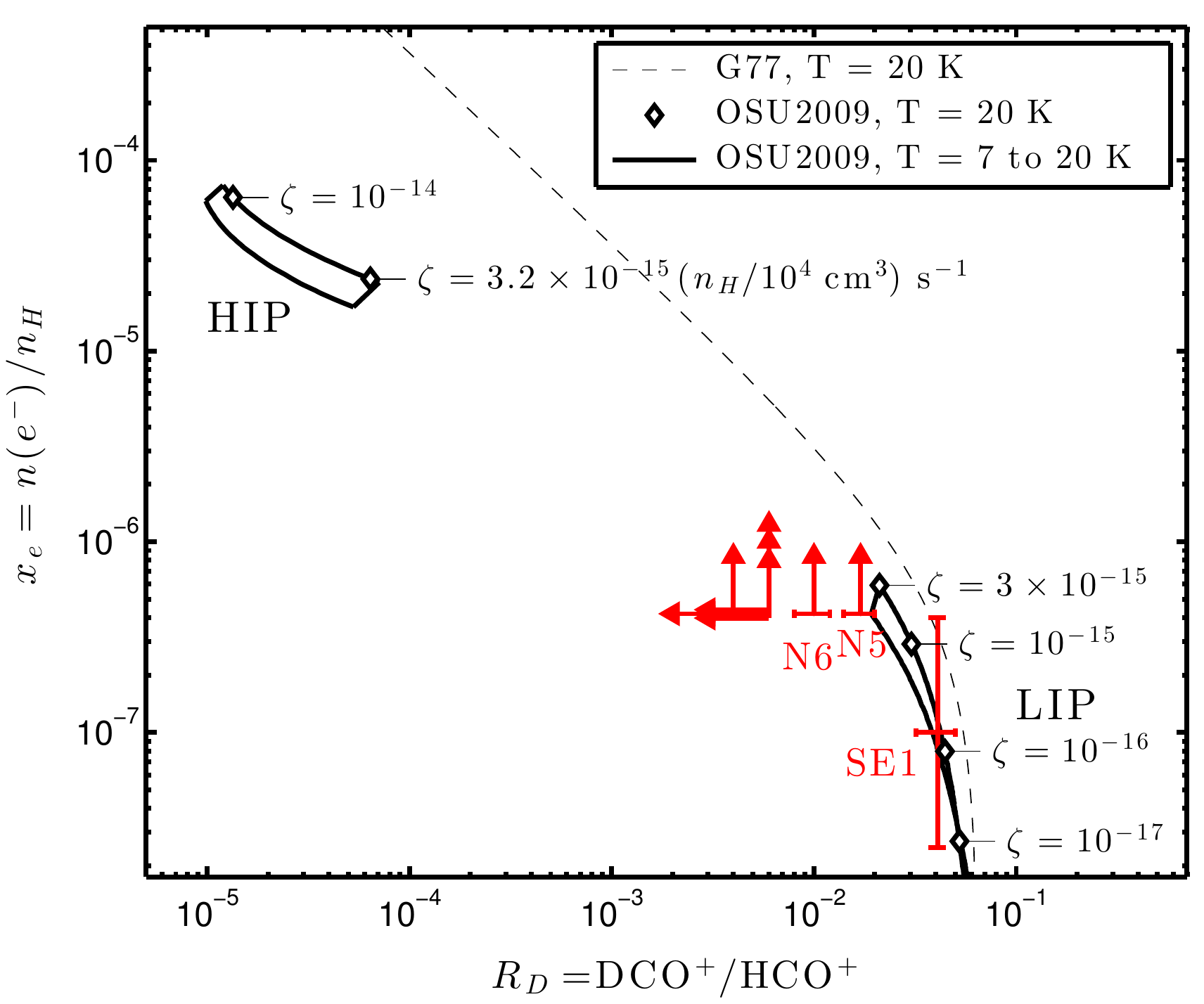}
  \caption{\label{fig:result} 
    $\xe=n(\e)/n_{\h}$ as a function of \rd. Values from the analytical method (G77) in \S~\ref{sec:analytical-method} are given at $T=20$~K (dashed line).
Results from our calculation (see \S~\ref{sec:numerical}) for temperatures between 7~K  and 20~K are contained within the solid lines. $\zeta$ values are given for $n_\h=10^{4}$~\cmmm, although the model only depends on the $\zeta/n_\h$ ratio (see text).
\rd\ values or upper limits  as derived from observations are {indicated by the red symbols; see \S \ref{sec:results-W28}}.
    \label{fig:xe_Rd}}
\end{figure}

Figure~\ref{fig:result} shows \xe\ as a function of \rd, as
predicted from  Eq.~\ref{eq:xe}, assuming a kinetic temperature of
20~K (blue dashed line). We note two regimes in the dependence of \xe\ on \rd.  {For}
low \rd\ values ($\lesssim 10^{-2}$), \xe\ is proportional to $1/\rd$
with a factor that depends on the chemical reaction rates and the \hd\
abundance. {For} higher \rd\ values, \xe\  drops sharply. In this
regime, \xe\ varies by more than two orders of magnitude
when \rd\ is changed by only a factor of two. The slope becomes
steeper with increasing temperature, due to the predominance of the
reverse reaction of the formation of \hhdp\
(Table~\ref{tab:reactions}). This indicates that accurate values of
\xe\ in dark clouds through this method require extremely accurate
values of \rd. {It also suggests} that in regions with higher
ionisation, where \xe\ is proportional to $1/\rd$, \dcop\ {will be} difficult to detect.

\subsection{Numerical models}
\label{sec:numerical}

To assess the validity of the analytical approach of G77 and C98, we
have solved  the OSU 2009\footnote{\url{http://www.physics.ohio-state.edu/~eric/research.html}}  chemical network for the abundances of \dcop, \hcop, and \e, using the \texttt{astrochem}\footnote{\url{http://smaret.github.io/astrochem/}} code.  The time
evolution of the gas-phase abundances was followed until a steady state
was reached, after typically 10~Myr. {The underlying hypothesis is that the cloud was already molecular when it was irradiated by the CRs emitted at the SN explosion. The chemical changes caused by the sudden CR irradiation are dominated by  ion-neutral reactions, whose timescale is only $\sim 10^2$~yr, much shorter than the W28 SNR age ($\sim 10^4$~yr; see Introduction).}
Details {of} the models are given in Appendix \ref{app:model}.
In one calculation, the gas is shielded by 20~mag of visual extinction, such that UV photons can safely be
ignored.

As anticipated from Eq.~\ref{eq:RH}, we
found that the abundances mostly depend on the $\zeta/\nh$ ratio,
rather than on $\zeta$ and \nh\ separately. This behaviour is similar
to photon-dominated regions in which a good parameter is the ratio of
the UV radiation field to the total density. For each temperature
from 7~K to 20~K, a series of calculations with $\zeta/\nh$ increasing
from \dix{-22}\cmmmp\s\ to \dix{-18}\cmmmp\s\ {were} performed, and the
steady-state values of \xe\ and \rd\ {were} recorded. 

{In these calculations, we assumed a standard CO abundance of $\sim 7.3\ 10^{-5}$ in the cloud. Indeed, since the density of the studied clouds is relatively low (Table~\ref{tab:results}), we do not expect CO depletion to play an important role here. However, in general, one has to take into account this uncertainty. In addition, our calculations do not consider separately the H2 ortho-to-para ratio, which is known to affect the DCO$^+$/HCO$^+$ abundance ratio when it is larger than about 0.1 \citep[e.g.][]{2011ApJ...739L..35P}. Our choice here is based on the published observations that indicate that the H2 ortho-to-para ratio is smaller than about 0.01 in molecular clouds \citep[e.g][] {2009A&A...506.1243T,2012A&A...537A..20D}.}


\subsection{A new view of the \dcop/\hcop\ method}

The results of the numerical models are shown in Fig.~\ref{fig:result}. As expected, the
ionisation fraction \xe\ increases with $\zeta/\nh$, with \rd\
decreasing in the process. There is good overall agreement between the
analytical and numerical predictions for $\rd \gtrsim 2\tdix{-2}$. In
this high-\rd\ regime, the small {differences} between analytical and
numerical values are due to the abundances of HD and CO not being
constant as originally assumed by G77 and C98. Instead, as $\zeta/\nh$ and  \xe\ 
increase, atomic deuterium becomes more
abundant. Similarly, the CO abundance decreases because of the
dissociating action of CRs. When \rd\ decreases and reaches
$\approx 2\tdix{-2}$, the abundances predicted by the numerical model
change dramatically to a regime characterised by large values of \xe\
and low values of \rd. This jump corresponds to the well-known
transition from the so-called low ionisation phase (LIP) to the high
ionisation phase (HIP) \citep{pineaudesforets1992, LeBourlot1993}, and
is due to the sensitivity of interstellar chemical networks to
ionisation. The LIP is associated with \rd\ larger than $10^{-2}$,
whilst the HIP is characterised by $\rd\lesssim 10^{-4}$.  In our
calculations, the LIP-HIP transition occurs at $\zeta/\nh \sim
3\times10^{-19}\,\cmmmp\s$, and we note that this value depends only
slightly on the temperature, although it is known to depend on other
parameters such as the gas-phase abundance of metals \citep{2006A&A...451..551W}. {A detailed
analysis of the LIP-HIP transition is, however, not the aim of this
study.}
Here, it is rather  the existence of this instability which is of interest since it produces  a sharp {difference} between the analytical
and {the} numerical predictions from the ionisation point of view. 
{Application to a practical case shows that what changes is not the jump itself but (slightly) the $\zeta$ at which it occurs \citep[see e.g.][]{CC2011}.}
In the
former, the variations of \xe\ and \rd\ are continuous and, as already
mentioned, predict $\xe \sim 1/\rd$ in the low-\rd\ regime. This
scaling is, however, not observed in the numerical models, and is replaced by a
discontinuous variation of both \xe\ and \rd. The present calculations
show that the LIP is characterised by $\rd=\dix{-2}-\dix{-1}$, $\xe
\lesssim 5\tdix{-7}$, and the HIP is characterised by $\rd\approx$ few
\dix{-5} and $\xe \approx$ few \dix{-5}. 

As shown in Fig.~\ref{fig:abundances}, the low values of \rd\ in
the HIP are due to a very low abundance of \dcop, whilst the abundance
of \hcop\ decreases by a smaller amount. This has important
consequences when using the \dcop/\hcop\ method to derive the
ionisation fraction and CR ionisation rate. First, it must be
recognised that this method may provide a value of \xe\ only for
LIP-dominated gas conditions. In other words, where \dcop\ is detected, the line of sight is dominated by low-\xe\ gas. For lines of sight dominated by HIP
gas, the abundance of \dcop\ is expected to be well below detectable
thresholds, such that only upper limits on \rd\ can be derived. Yet,
an upper limit {on} \rd\ still provides essential information, since it
is associated with a lower limit {on} \xe, which in turn corresponds to
a lower limit of $\zeta/\nh$. On the contrary, for LIP-dominated lines
of sight, the value of \xe\ and $\zeta/\nh$ may be derived directly
from \rd, although \xe\ is extremely sensitive to uncertainties on $R_D$ in this regime.

%
\section{The CR ionisation rate in W28}
%
\label{sec:results_CRI}
\subsection{Constant density and temperature cloud analysis}
\label{sec:results-W28}
A new view of the \dcop/\hcop\ method thus emerges, which stresses its
strengths and limitations. {The} method allows the determination of
the ionisation fraction  \xe\ and  the $\zeta/\nh$ ratio for gas in the LIP, and
provides lower limits of \xe\ and $\zeta/\nh$ for gas in the HIP. 
In the following, we apply this method to the sample of observed points, using
the constraints on the gas temperature and density, and the \rd\ value
in each point (Table~\ref{tab:results}).
We emphasise that the model calculations summarised
  in Fig.~\ref{fig:xe_Rd} assumed constant density and gas
  temperature. In the next section, we will discuss how {the
  \dcop/\hcop\ method} can be used {to
  constrain} \xe\ and $\zeta/\nh$, taking into account the thermal
structure of the cloud.

{Of the 16 lines of sight initially observed in CO, 12
  were also detected  in \htrcop, of which 4 led to \rd\ determinations
  and 5 to upper limits (Table~\ref{tab:results}).}
The four points with measured \rd\ are N5, N6, SE1, and N2. In the
following analysis, we exclude N2 as it coincides with a protostar,
which means that a more accurate analysis {taking} into account the
structure of the protostar and the inner ionisation is necessary.
The values obtained towards N5, N6, and SE1 are shown in
Fig.~\ref{fig:result}. 

\begin{figure}
  \centering
  \includegraphics[width=\hsize]{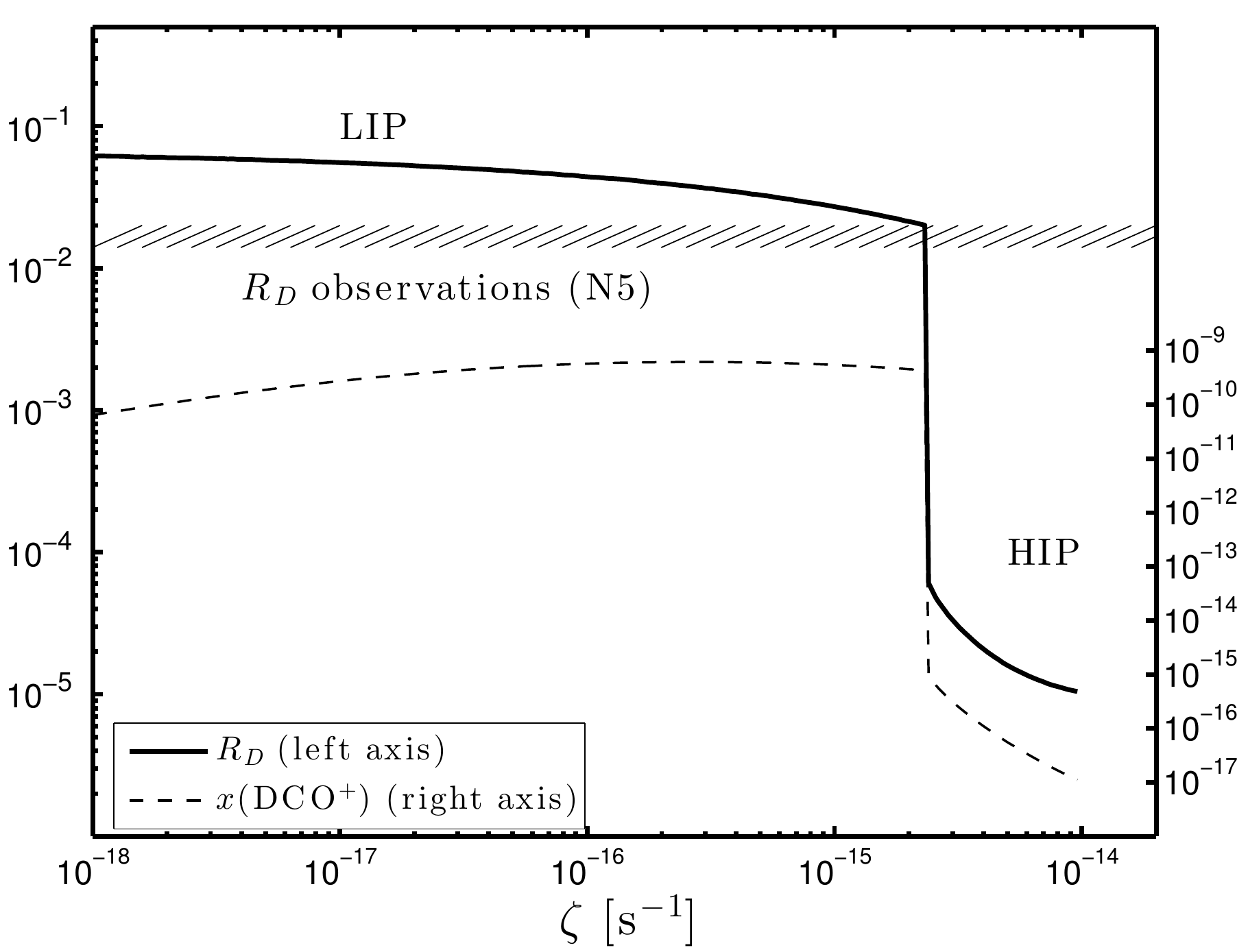}
  \caption{\label{fig:abundances} $\rd=\dcop/\hcop$ (thick line, left axis) and $x(\dcop)=n(\dcop)/n_\h$ (dashed line, right axis) as a function of $\zeta$, for
    $T=10$~K and $n_\h = 8\ 10^3$~\cmmm, i.e. {the physical conditions characterising} position N5. The HIP and LIP are
    marked. The hatched area shows the range of observed \rd\ at that
position.}
\end{figure}

The SE1 point lies on the LIP branch, enabling a determination of the ionisation
fraction $\xe$=(0.15--4)$\tdix{-7}$ and of the CR ionisation rate
$\zeta$=(0.2--20)$\tdix{-17}$~\s.
On the contrary, the values of \rd\ towards N5 and N6 lie in the gap
between the LIP and HIP branches, even when considering a kinetic
temperature as high as 20~K, temperature which is larger than {the values derived} for
these positions. In these cases, adopting
$n_\h=2n(\hh)\gtrsim4\tdix{3}$~\cmmm\ (Table~\ref{tab:results}),
Fig.~\ref{fig:xe_Rd} provides the following lower limits: $\xe \gtrsim
4\tdix{-7}$ and $\zeta \gtrsim 1.3\tdix{-15}$ \s\, {for} both
points. We note that the detection of \dcop\
  indicates that the line of sight includes a non-negligible amount of
  LIP, which can serve to further constrain the value of $\zeta/\nh$.
This can be seen in Fig.~\ref{fig:abundances}, which
shows \rd\ and \dcop\ as a function of $\zeta$, for a density and
temperature appropriate to position N5 (Table~\ref{tab:results}). The
measured \rd\ intersects the model predictions at the edge of the
LIP/HIP jump, at $\zeta \approx 2.5\times 10^{-15}$~\s. More
importantly, the figure shows that the gas is neither entirely in the
LIP nor HIP state as expected from the detection of \dcop.
A similar plot has also been obtained  for N6, leading to the same
conclusion.

Finally, the non-detection of \dcop\ {in the other lines of sight}
 leads to upper limits on \rd\ that are well
outside the LIP branch. At these positions, the gas is very likely to be almost
entirely in the HIP state, which means that $\zeta/\nh \gtrsim
3\tdix{-19}$~\cmmmp~\s. {An exception is} the point SW4, where an
extremely energetic outflow has been detected \citep{1988A&A...197L..19H}, and where the
\hcop\ is therefore likely contaminated by the outflowing material.

\subsection{Constant density cloud analysis}

As discussed in the previous section, the points N5 and N6 are likely
composed of a {mixture} of gas in the LIP and HIP state. This is similar to
the situation observed in W51C-E (CC2011). In that case, $\zeta$ was
estimated with a model that takes into account the thermal and
chemical structure of a constant density cloud, where a fraction (the
deepest) is in the LIP and the rest in the HIP (Fig.~2 in
CC2011). Here, we do a similar analysis, using basic arguments instead
of a sophisticated model, and we show that it leads to similar
results, namely the determination of $\zeta$ to within a factor of 2. The
advantage of this analysis is that it shows in a straightforward way
the uncertainty due to the model parameters.

The crucial point is understanding what causes the gas to flip from
the HIP to the LIP state going deeper into the cloud.
Since the column density is too low to appreciably reduce $\zeta$
across the cloud, the only macroscopic quantity that changes is the
gas temperature. Specifically, the temperature increases {by a few~K} (in the
UV-shielded region) going deeper into the cloud because the CO line
opacity increases and, consequently, the line cooling becomes less
efficient. The effect is larger for larger $\zeta$ as the heating,
dominated by the CR ionisation, is less compensated by the line
cooling.

\begin{figure}
  \centering
  \includegraphics[width=\hsize]{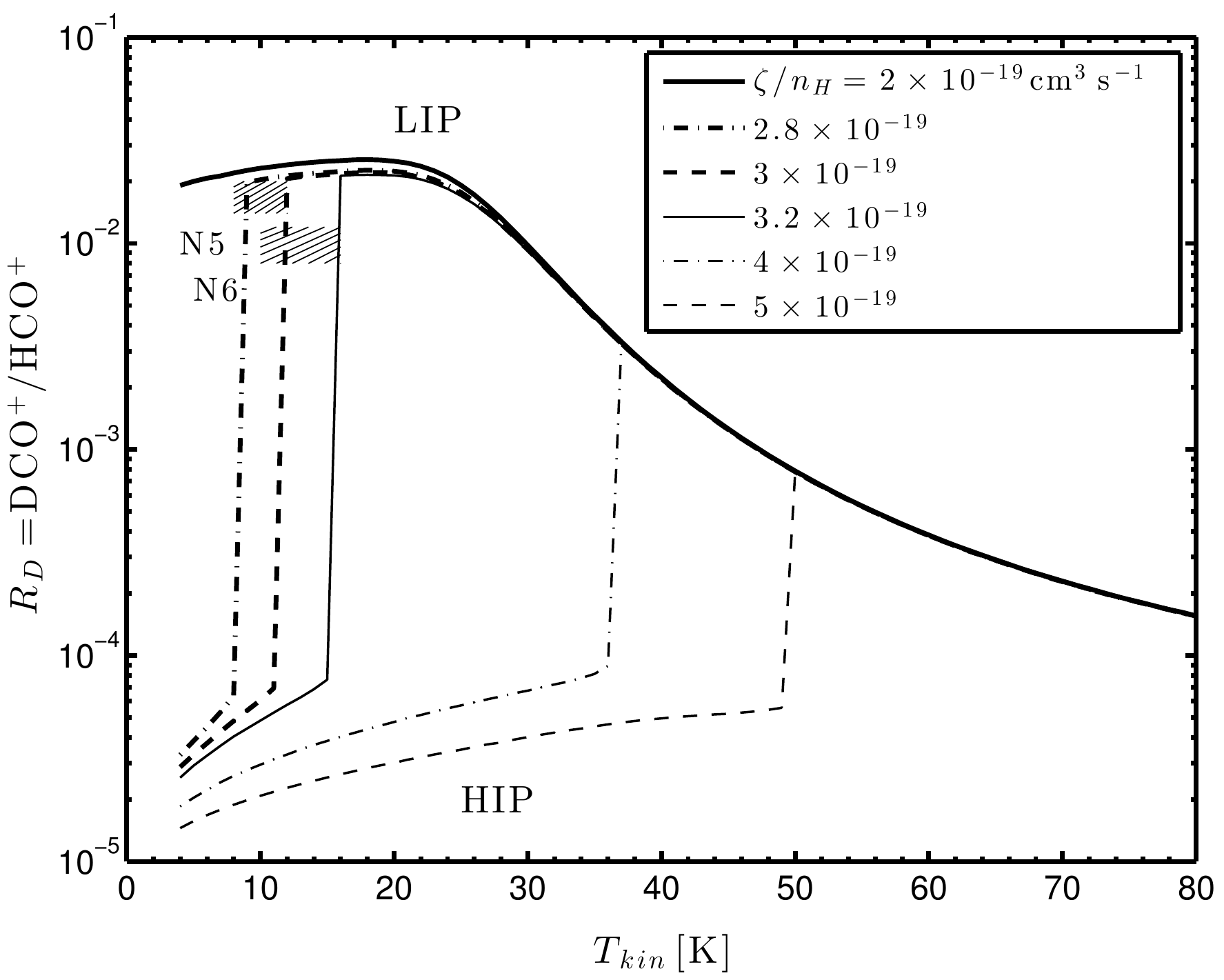}
  \caption{\label{fig:RD_T} \rd\ as a function of the gas temperature
    $T_{kin}$ for different values of $\zeta/\nh$: from 2 to 5
    $\times10^{-19}$ \s, as marked. Note that for $\zeta/n_\h \leq
    2\times 10^{-19}$ cm$^3$  \s (thick solid line), the cloud is always in the
    LIP, regardless of the temperature.  For $\zeta/n_\h > 5\times
    10^{-19}$ cm$^3$ \s (thin dashed curve), the cloud is always in the HIP for
    temperatures $\leq$ 50 K. Hatched areas show  observations of N5 and N6. We {assume} $A_V=20$~mag.}
\end{figure}

It is instructive to see how the \rd\ ratio changes as a function of
the gas temperature for different $\zeta$. This is shown in
Fig.~\ref{fig:RD_T}, for a range of temperatures (5--80 K) and
$\zeta/\nh$ {(2--5$\tdix{-19}$ cm$^3$ \s, appropriate to the N5 and
N6 points)}. 
In these calculations, we consider a cell of gas of constant density,
shielded by 20 magnitudes of visual extinctions as before, such that
the ionisation is driven by the CRs.
The figure shows important features:\\
i) for $\zeta/\nh \lesssim 2\tdix{-19}$ cm$^3$ \s, the cloud is always
in the LIP, regardless of the temperature;\\
ii) for $\zeta/\nh \gtrsim 5\tdix{-19}$ cm$^3$ \s, the cloud is
always
in the HIP for temperatures lower than 50 K;\\
iii) for intermediate values of $\zeta/\nh$, the gas flips from HIP
to LIP with increasing temperature, and the larger $\zeta$ is, the larger
the temperature where the flip occurs.
%


{These calculations show that there is a range of ionisation rates in which the gas is extremely sensitive to temperature variations. A tiny increase in temperature is sufficient to make the gas flip from the HIP to the LIP. In particular, for position N5, there is such a combination of values of $R_D$,  $T_{\rm kin}$, and $\zeta/n_\h$ (see Fig.~\ref{fig:RD_T}) that places it precisely in a region where the transition from HIP to LIP can be triggered by an increase in $T_{\rm kin}$ as small as a few~K. A similar argument applies to N6.}
In addition, in regions exposed to an
enhanced CR ionisation rate, the outer part characterised by large
ionisation fractions will be extended farther into the cloud,
thus decreasing the relative amount of LIP with respect to the HIP.

Based on the derived kinetic temperatures and values of $R_D$
(Table~\ref{tab:results}) and using Fig.~\ref{fig:RD_T}, we can
further constrain the value of $\zeta/\nh$.  Towards N5, the
temperature was found to be $10\pm2$~K, while $R_D=0.014-0.020$. When
{inserted into} Fig.~\ref{fig:RD_T}, these delineate a region that is compatible with a
narrow range of $\zeta/\nh=(2.8-3.0)\tdix{-19}$\cmmmp~\s. For the N6
line of sight, we find similar values, $(2.9-3.2)\tdix{-19}$~\cmmmp~\s. The
densities derived from the analysis and their uncertainties then lead
to cosmic ray ionisation rates of $(1.3-3.3)\tdix{-15}$~\s\ and
$(1.3-4.0)\tdix{-15}$~\s\ for N5 and N6 respectively. Results are
summarised in Table~\ref{tab:results} and  are included in
Figs.~\ref{fig:Padovani} and~\ref{fig:lowlim} 

%
\section{Discussion}
\label{sec:discussion}
%
\begin{figure}
\centering
\includegraphics[width=\hsize]{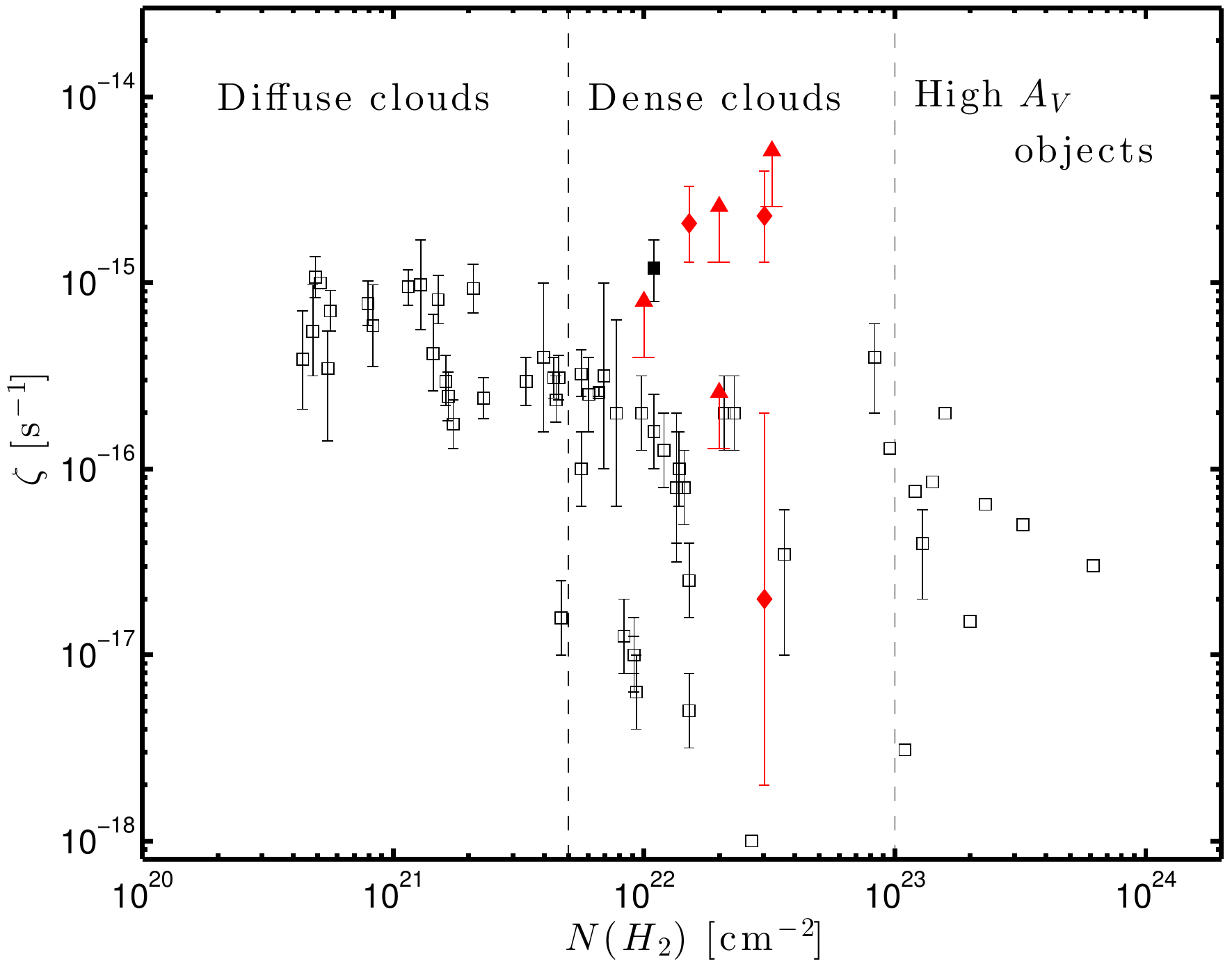}
\caption{\label{fig:Padovani} Compilation of measured $\zeta$ in
  different objects (open squares), as reported by
  \cite{2013ASSP...34...61P}. The black filled square {denotes} W51 \citep{CC2011}. Red points and lower limits report the
  values derived in this work. The dashed lines show the range of
  column densities $(0.5-10)\times10^{22}$~\cmm, typical of dense molecular clouds, corresponding to visual extinctions of 5 and 100~mag, respectively. On the left lie
  the diffuse clouds and on the right highly obscured environments such as infrared dark clouds or protoplanetary discs.}
\end{figure}
Table~\ref{tab:results} lists the observed positions and the
corresponding CR ionisation rates derived {using} the method
described in the previous section. {With}
the exception of the SE1 point, in all other points $\zeta$ is at
least 10 to 260 times larger than the standard value
($1\times10^{-17}$~\s) in Galactic clouds. This is  shown in
Fig.~\ref{fig:Padovani}, where we {present} a compilation of the $\zeta$
measured in various objects \cite[from][]{2013ASSP...34...61P}, plus
our measurements.  In the range of column densities $(0.5-10) \times
10^{22}$ cm$^{-2}$, typical of dense molecular clouds, the points in
which we derived $\zeta$ are those with the highest values, together
with the CC2011 point (filled square).  The first conclusion of this work is,
therefore, that clouds next to SNR are indeed irradiated by an
enhanced flux of CRs of relatively low energy (see below for a more
quantitative statement on the CR particle energies).

\begin{figure}
  \centering
  \includegraphics[width=\hsize]{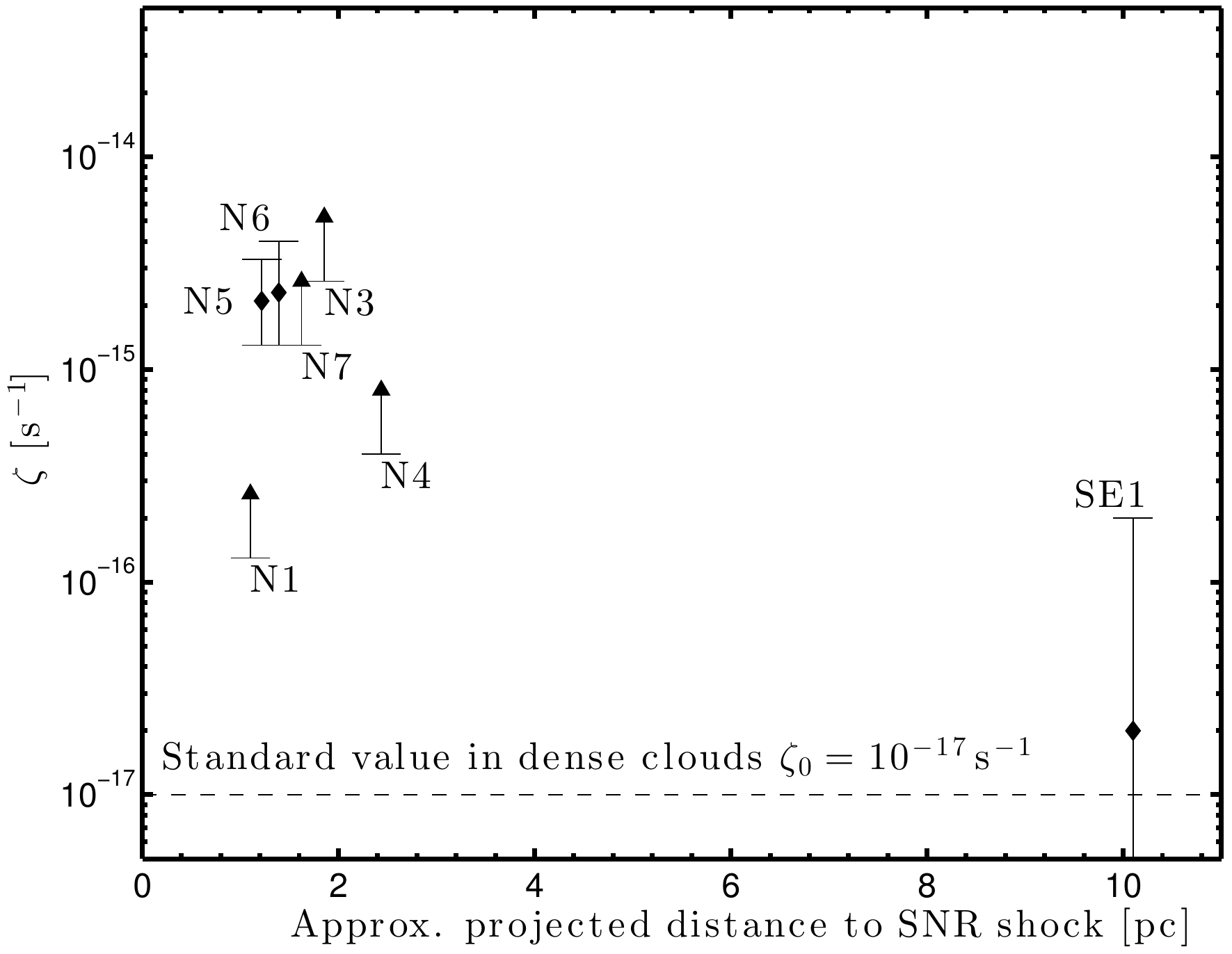}
  \caption{\label{fig:lowlim} CR ionisation rate $\zeta$ as a function
    of the approximate projected distance from the SNR radio boundary
    (blue circle in Fig.~\ref{fig:map}), assuming a W28 distance of
    2~pkc. We note that the $\zeta$ error bars are dominated by the
    uncertainties on the \hh\ densities (see text). }
\end{figure}
%
{The} dependence of $\zeta$ on the projected distance from
the SNR radio boundary (assuming a W28 distance of $2$~kpc) {is shown in Fig.~\ref{fig:lowlim}}.
Remarkably, the point farthest ($\sim10$~pc) from the SNR edge is the
one with the lowest $\zeta$.  Actually, it is the only point where the
gas is predominantly in the LIP state. All other points,
at distances $\lesssim 3$~pc, have at least a fraction of the gas in
the HIP, namely they have a larger \xe\ and $\zeta$.  Of
course, this analysis does not take into account the 3D structure of
the SNR complex. Yet, this can still provide us {with constraints} on the propagation properties of CRs, {as will} be
discussed in the following.

{Valuable additional information} is provided {by observations} in
the \ensuremath{\gamma}-ray\ domain. Both the northern and southern clouds coincide with
sources of TeV emission, as seen by HESS. This means that the clouds
are illuminated by very high energy ($\gtrsim10$~TeV) CRs, which already
escaped the SNR expanding shell and travelled the $\gtrsim 10$~pc (or
more, if projection effects play a role) to the southern
cloud. Conversely, the low CR ionisation rate measured in SE1 tells us
that the ionising lower energy CRs remain confined closer to the SNR.
In the same vein, GeV emission has been detected towards the northern
region but only towards a part of the southern one. This difference
between the GeV and TeV \ensuremath{\gamma}-ray\ morphology has been interpreted as a
projection effect: the {portion} of the southern region that exhibits a lack
of GeV emission is probably located at a distance from the shock
significantly larger than the projected one, $>10$~pc, and thus can be
reached by $\gtrsim$~TeV CRs but not by $\gtrsim$~GeV ones
\citep{2010sf2a.conf..313G,2010MNRAS.409L..35L,2013MNRAS.429.1643N}.
Remarkably, the SE1 point is located in the region where the lack of
GeV emission is observed.

The picture that emerges is that of a stratified structure with CRs of
larger and larger energies occupying larger and larger volumes ahead
of the shock.  Within this framework, it is possible to estimate the
CR diffusion coefficient in the region. This can be done by recalling
that in a given time $t$, CRs diffuse over a distance $R \sim \sqrt{D
  \times t}$, where $D$ is the energy dependent diffusion
coefficient. For the situation under {examination}, one gets
\begin{equation}
\label{eq:DGeV}
D_{(\approx 10~{\rm GeV})} \gtrsim 3 \times 10^{27} \left( \frac{R}{10~{\rm pc}} \right)^2 \left( \frac{t}{10^4~{\rm yr}} \right)^{-1} {\rm cm^2/s}\ ,
\end{equation}
where $D_{(\approx 10~{\rm GeV})}$ is the diffusion coefficient of
$\approx 10$~GeV CRs, which are those responsible for the $\approx$~GeV
\ensuremath{\gamma}-ray\ emission, and $t$ is the time elapsed since the escape of CRs
from the SNR. The value obtained in Eq.~\ref{eq:DGeV} is in
substantial agreement with more accurate studies \cite[see
e.g.][]{2013MNRAS.429.1643N}.

The diffusion coefficient obtained in Eq.~\ref{eq:DGeV} can  then be
rescaled to lower energies, according to $D \propto p^s \beta$, where
$p$ is the particle momentum, $\beta = v/c$ its velocity in units of
the speed of light, and $s$ depends on the spectrum of the ambient
magnetic turbulence. The typical value of $s$ in the interstellar
medium is poorly constrained to be in the range 0.3 to 0.7
 \citep{2011arXiv1110.2981C}. In the following, we adopt $s = 0.5$. To estimate
the diffusion length of low energy CRs, one has to keep in mind that,
while CRs with energies above $\approx$~GeV are virtually free from
energy losses (the energy loss time for proton--proton interactions in
a density $\nh \approx 10^3$~cm$^{-3}$ is comparable to the age of the
SNR), lower energy CRs suffer severe ionisation losses over {a} short
timescale \citep{1990acr..book.....B}:
\begin{equation}
\tau_{ion} ~\approx~ 14 ~ \left( \frac{\nh}{10^3~{\rm cm}^{-3}} \right)^{-1} \left( \frac{E}{\rm MeV} \right)^{3/2} {\rm yr} ~ .
\end{equation}
This approximate expression is sufficiently accurate in the range of
energies spanning  1-100 MeV.  The diffusion length of low
energy CRs can then be estimated by equating the diffusion time $\tau_d
\sim R_d^2/D$ to the energy loss time $\tau_{ion}$, which gives $R_d
\approx$~0.02, 0.3, and 3~pc for CRs of energy 1, 10, and 100 MeV,
respectively. This implies that only CRs with energies $\gtrsim
100$~MeV can escape the shock and spread over a distance of
3~pc or more, and thus these are the CRs that play a major
role in ionising the gas.  Whether the ionisation of the gas is due
directly to these CRs or to the products of their interaction with the
gas (namely slowed down lower energy CRs) remains an open question. It
is remarkable that the particle energies of ionising CRs ($\approx
0.1-1$~GeV) also make them  capable of producing sub--GeV \ensuremath{\gamma} rays,
given that the {kinetic} energy threshold for $\pi^0$ production is
$\approx 280$~MeV.

Of course, the order of magnitude estimates discussed in this section cannot substitute in any way more sophisticated calculations, yet they clearly indicate an intriguing possible link between low and high energy observations of SNR environments. 
In fact, in the scenario described above, the very same CRs are responsible for both ionisation of the gas and production of low energy \ensuremath{\gamma} rays. 
If confirmed, such a link would constitute robust evidence for the presence of accelerated protons in the environment of the SNR W28, a thing that would bring further support to the idea that SNR are the sources of Galactic CRs.
Additional theoretical investigations are needed in order to examine and possibly rule out alternative scenarios which may include other contributions to the ionisation rate (e.g. CR electrons, X-ray photons) or different means of propagation (e.g. straight--line or advective propagation of low energy CRs).

\section{Conclusion}\label{sec:conclusion}

In this work, we presented
  new observations to measure the CR ionisation rate in molecular
  clouds close to {supernova remnants (SNR)}. In doing so, the \dcop/\hcop\ method was also
  revisited. The major results may be summarised as follows.

\noindent
 1) We observed the {two lowest rotational
  transitions of} \thco\ and \cdo\  towards 16 positions in the northern and southern clouds
  close to the SNR W28. The four lines were detected in emission
  towards 12 of these positions, where we could, therefore, derive the
  physical conditions using a non-LTE LVG analysis. With the exception
  of one position (N2) coinciding with a protostar in the region, we
  derived \hh\ densities and temperatures typical of molecular clouds,
  namely $n_\hh = (0.2 - 10) \times 10^3$~\cmmm\ and $T = 6 -
  24$~K. We searched for \htrcop\ and \dcop\ line emission in the above
  12 positions, and detected it in 9 and 4, respectively.  {From these data},
  we could derive the $R_D = \dcop/\hcop$ in 4 positions, one of
  which coincides with the {protostar}, and give 
  upper limits {for} the remaining 5 positions.\\
\noindent 
2) We reinvestigated the \dcop/\hcop\ method used to derive the
  ionisation fraction $\xe=n(\e)/\nh$ and the relevant CR ionisation
  rate $\zeta$ causing it. {To this aim, we compared the
    steady-state abundances of \hcop, \dcop, and \e\ as
    predicted by the analytical model of G77, to numerical
    calculations, assuming constant density and gas temperature. The
    numerical model leads to two well separated regimes of ionisation,
    also known as the low- and high-ionisation phases (LIP and HIP)
    \citep{pineaudesforets1992}. In the context of this work, these
    two phases lead to two separated regimes in terms of $\zeta$ and \rd\ values:\\
     i) for $\zeta/\nh
    \lesssim 3\times 10^{-19}$~\cmmmp~\s, the gas is in the LIP, where
    $R_D \gtrsim 2\times 10^{-2}$ and $\xe \lesssim6\times 10^{-7}$.
    In this regime, the {dependence of \xe\ on $R_D$} is very steep
    leading to large uncertainties on \xe;\\
     ii) for $\zeta/\nh
    \gtrsim 3\times 10^{-19}$~\cmmmp~\s, the gas is in the HIP, where
    $R_D \lesssim 10^{-4}$ and $\xe \gtrsim 2 \times 10^{-5}$. In this
    regime, \dcop\ is not detectable and the numerical prediction for
    \xe\ differs significantly from the analytical one.}

\noindent
Therefore, the \dcop/\hcop\ abundance ratio can provide a measure of
\xe\ and $\zeta$ in the LIP, and only lower limits if the gas
is in the HIP.

\noindent
3) We found only one position, SE1, in the LIP, where $R_D= 0.032 -
0.05$, $\xe=(0.3-4) \times 10^{-7}$ and $\zeta=(0.2-20)\times
10^{-17}$~\s. Two positions, N5 and N6, lie in the gap between the LIP
and HIP, namely the gas is neither entirely in the LIP nor in the HIP,
although it certainly contains a fraction of gas in the LIP, where
\dcop\ is detectable (and detected). {The
  jump from the HIP to the LIP when penetrating farther into the
  cloud is associated {with} an increase in the temperature, and we showed
  that model calculations at several temperatures further constrain
  the value of $\zeta$.}  The uncertainty
in $\zeta$ towards these positions is dominated by the uncertainty in
the \hh\ density {and the derived values are}
$\zeta=(1.3-3.3)$ and $(1.3-4.0)$ $\times 10^{-15}$~\s\ for N5 and N6,
respectively.  Towards the remaining 5 positions with upper limits
{on} $R_D$, the derived $\zeta$ values
are at least 10 to 260 times higher than the standard value
of $1\times10^{-17}$~\s.

\noindent
4) {The points of the northern cloud have the largest CR
  ionisation rates measured so far in the Galaxy. The point towards the southern cloud
  is, on the contrary, consistent with the average galactic CR
  ionisation rate of molecular clouds not interacting with a SNR.} Since the northern and
southern clouds have projected distances from the SNR shock of $\leq3$
and $\sim10$~pc, respectively, this can be explained by the fact that
the low energy ionising CRs have not reached the southern cloud yet. On
the other hand, the observations show that both the northern and
southern clouds coincide with TeV emission sources, suggesting that
high $\gtrsim10$~TeV CRs have reached both. {This is
  also consistent with} \ensuremath{\gamma}-ray\ emission sources
{coinciding} with the northern cloud but only partially
with the southern cloud, indicating that the former is irradiated by
$\approx0.1-1$~GeV CRs, while only the nearest
{portion} of the southern cloud is {so affected}.

\noindent
5) {The emerging picture is that of energy-dependent diffusion
  properties of hadronic CRs. The high-energy CRs responsible for TeV
  \ensuremath{\gamma}-ray\ emission through $\pi^0$-decay can diffuse far ahead of the SNR
  shock, while the low-energy CRs ($0.1-1$~GeV), responsible for both
  the low \ensuremath{\gamma}-ray\ emission and the ionisation of the gas, remain closer
  to the SNR shock. The present work thus gives  first
  observational evidence to the theoretical predictions that hadrons
  of energy $0.1-1$~GeV contribute most to the ionisation in dense gas
  \citep{padovani2009}.}

\begin{acknowledgements}
We warmly thank Marco Padovani for providing us with his compilation and for useful discussions.
  This work has been financially supported by the Programme National
  Hautes Energies (PNHE).  Based on observations carried out with the
  IRAM 30m telescope. IRAM is supported by INSU/CNRS (France), MPG
  (Germany) and IGN (Spain).  S.Gabici acknowledges the financial
  support of the UnivEarthS Labex Program at Sorbonne Paris Cit\'e
  (ANR-10-LABX-0023 and ANR-11-IDEX-0005-02).
\end{acknowledgements}


\nocite{*}
\bibliographystyle{aa} 
\bibliography{vaupre14} 

\appendix
\section{Chemical models}
\label{app:model}

We solved the OSU
2009\footnote{\url{http://www.physics.ohio-state.edu/~eric/research.html}}
chemical network using the
\texttt{astrochem}\footnote{\url{http://smaret.github.io/astrochem/}}
code.

\texttt{astrochem} is a numerical code
that computes the time-dependent chemical abundances
in a cell of gas shielded by a given visual extinction $A_V$
  and with given physical parameters: the total H density $n_\h$, the
gas kinetic temperature $T_{\rm kin}$, and the dust temperature
$T_{\rm d}$. It also takes as an input the initial chemical
abundances and the CR ionisation rate.  We followed the
  abundance until a steady state was reached, for a grid of
models covering a large range of physical conditions
$(n_\h,T_{\rm kin})$ and CR ionisation rates $\zeta$, at a given
$A_V=20$~mag, far inside the cloud, where the gas is shielded from the
UV radiation field and the ionisation is dominated by CRs. The results
only depend on the $\zeta/n_\h$ ratio (see
\S~\ref{sec:numerical}). {The role of the dust in \texttt{astrochem} is limited to the absorption and desorption processes, namely no grain surface chemistry is considered. As discussed in Sect.~\ref{sec:numerical}, neither process is relevant to the present discussion.}
Initial conditions were taken from the low-metal abundances as
in \cite{1982ApJS...48..321G} and \cite{2006A&A...459..813W} using an
updated He/H relative abundance of 0.09 \citep{Asplund:2009eu}.  In
Table~\ref{tab:CI}, we list the range of physical parameters used in
this study.

The OSU network contains 6046 reactions involving 468 species. We
appended 12 reactions labeled \ref{kf}-\ref{kcop} in
Table~\ref{tab:reactions} involving the deuterated species: D, HD,
\hhdp, and \dcop. Chemical rates are taken from
\cite{2000A&A...361..388R}.

\begin{table}
  \caption{Reduced chemical network for the analytical derivation of \dcop/\hcop. \label{tab:reactions}}
  {\centering
    \resizebox{ \hsize}{!}{
      \begin{network}
        \toprule \midrule 
        \multicolumn{4}{c}{Reaction}&Reaction rate [\cmmmp.s$^{-1}$] \\ 
        \midrule
        \label{zeta}&$\text{CR}+\hh$&$\carrow{\zeta}$& $\hhp + \e$ &$\zeta$ [\s]\\
        \label{khhp}&\hhp + \hh &$\carrow{k_{\hhp}}$&$\htrp + \h $ &$k_{\hhp}  = 2.1 \ 10^{-9} $\\
        \label{kh}&$\htrp +\co $&$ \carrow{k_H}$&$  \hcop + \hh$&$k_{H}=1.61\ 10^{-9}$\\
        \label{betaph}&\hcop$+\e$&$\carrow{\beta'}$&\co +\h&$\beta'=2.8\ 10^{-7}\left(\frac{T}{300}\right)^{-0.69}$\\
        \label{beta}&\htrp$+\e$&$\carrow{\beta}$&\h + \h + \h &$\beta=4.36\ 10^{-8}\left(\frac{T}{300}\right)^{-0.52}$\\
        \multicolumn{1}{l}{}&&&\hh +\h& $ \ \ \ \ \  + 2.34\ 10^{-8}\left(\frac{T}{300}\right)^{-0.52}$ \\
        \label{kp}&$\h +\h $ &$\carrow{k'}$& \hh & $k' = 4.95\ 10^{-17}\left(\frac{T}{300}\right)^{0.50}$\\
        \label{kf}&$\htrp + \hd $&$ \ceq{k_f}{k_f^{-1}}$&$  \hhdp + \hh$&$k_{f}=1.7\ 10^{-9}$\\
        \multicolumn{1}{l}{}&&&&$k_{f}^{-1}=k_{f}\ \text{e}^{-220/T}$\\
        \label{kd}&$\hhdp + \co$&$  \carrow{k_D}$&$  \dcop + \hh $&$k_D=5.37\ 10^{-10}$\\
        \label{betapd}&\dcop$+\e$&$\carrow{\beta'}$&\co +\d&$\beta'=2.8\ 10^{-7}\left(\frac{T}{300}\right)^{-0.69}$\\
        \label{ke}&$\hhdp + \e$&$ \carrow{k_e}$&$ \h+\h+\d$& $k_{e}=4.38\ 10^{-8}\left(\frac{T}{300}\right)^{-0.50}$\\ 
        \multicolumn{1}{l}{}&&&$\hh+\d$&$\ \ \ \ \ \ + 1.20\ 10^{-8}\left(\frac{T}{300}\right)^{-0.50}$\\
        \multicolumn{1}{l}{}&&&$\hd+\h$&$\ \ \ \ \ \  + 4.20\ 10^{-9}\left(\frac{T}{300}\right)^{-0.50}$\\
        \label{kpp}&$\h +\d $ &$\carrow{k''}$& \hd & $k'' = \sqrt{2} k'$\\
        \label{kdp}&$\hhdp + \co$&$  \carrow{k_D'}$&$  \hcop + \hh $&$k_D'=1.1\ 10^{-9}$\\
        \label{kfp}&$\htrp + \d $&$ \ceq{k_f'}{k_f'^{-1}}$&$  \hhdp + \h$&$k_{f}'=1.0\ 10^{-9}$\\
        \multicolumn{1}{l}{}&&&&$k_{f}'^{-1}=k_f'\ \text{e}^{-632/T}$\\
        \label{kcop}&$\co^+ + \hd$&$  \carrow{k_{\co^+}}$&$\dcop + \h$&$k_{\co^+}=7.5\ 10^{-10}$\\
        \bottomrule
      \end{network}
    }
    \\ }
  \begin{mytablecaption}
    The reduced network corresponds to the original description by
    \cite{Guelin1977} and \cite{Caselli1998}.   The rates of reactions \ref{zeta}-\ref{kp} are contained in the original OSU 2009 network. We
    appended deuterated reactions \ref{kf}-\ref{kcop} for
    which chemical rates are taken from \cite{2000A&A...361..388R}.
    Reaction \ref{kcop} is only dominant in the HIP and is not
    involved in the analytical determination of \dcop/\hcop.
  \end{mytablecaption}
\end{table}

\begin{table}
  \caption{Range of initial physical parameters  used in the
    \texttt{astrochem} code.\label{tab:CI}}
  {\centering
    \begin{tabular}{cc}
      \toprule \midrule
      Parameter&Range\\
      \midrule
      $A_V$&20 mag\\
      $n_\h$&$10^3$ to $10^4$ \cmmm\\
      $T_{kin}$&5 to 80 K\\
      $T_d$&20 K\\
      $\zeta$&$10^{-18}$ to $10^{-14}$ \s \\
      \bottomrule
    \end{tabular}\\}
\end{table}

\end{document}